\documentclass[preprint,aps,floats,nofootinbib,showpacs]{revtex4}
\usepackage{epsfig,epsf}
\usepackage{graphicx}
\begin{document}
\thispagestyle{empty}

\title{Neutron Stars for Undergraduates}
\author{
Richard R.\ Silbar and
Sanjay Reddy}

\affiliation{
Theoretical Division, Los Alamos National Laboratory, \\
Los Alamos, NM 87545}




\begin{abstract}

Calculating the structure of white dwarf and neutron
stars would be a suitable topic for an undergraduate thesis
or an advanced special topics or independent study course.  
The subject is rich in many different areas of physics accessible to a
junior or senior physics major, ranging from thermodynamics to quantum 
statistics to nuclear physics to special and general relativity. 
The computations for solving the coupled structure differential equations 
(both Newtonian and general relativistic) can be done using a symbolic 
computational package, such as Mathematica.
In doing so, the student will develop computational skills and learn 
how to deal with dimensions.
Along the way he or she will also have learned 
some of the physics of equations of state and of degenerate stars.

\end{abstract} 


\maketitle


\section{Introduction}

In 1967 Jocelyn Bell, a graduate student, along with her thesis 
advisor, Anthony Hewish, discovered the first pulsar, something from 
outer space that emits very regular pulses of radio energy.  
After recognizing that these pulse trains were so unvarying
that they would not support an origin from LGM's (Little Green
Men), it soon became generally accepted that the pulsar was 
due to radio emission from a rapidly rotating neutron star \cite{Landau}
endowed with a very strong magnetic field.
By now more than a thousand pulsars have been catalogued \cite{Princeton}.
Pulsars are by themselves quite interesting \cite{Jodrell}, but perhaps
more so is the structure of the underlying neutron star.
This paper discusses a student project dealing with that structure.

While still at MIT before coming to Los Alamos, one of us (Reddy) 
had the pleasure of acting as mentor for a bright British high school 
student, Aiden J.\ Parker.
Ms.\ Parker was spending the summer of 2002 at MIT as a participant in a 
special research program (RSI).
With minimal guidance she was able to write a Fortran program for solving the
Tolman-Oppenheimer-Volkov (TOV) equations \cite{TOV} to calculate masses and
radii of neutron stars (!).

In discussing this impressive performance after Reddy's arrival at LANL,
the question came up of whether it would have been possible (and easier)
for her to have done the computation using Mathematica (or some other
symbolic and numerical manipulation package).  
This was taken up as a challenge by the other of us, who also figured
it would be a good opportunity to learn how these kinds of stellar structure
calculations are actually done.  
(Silbar's only previous experience in this field
of physics consisted of having read, with some care, the chapter
on stellar equilibrium and collapse in Weinberg's treatise on gravitation 
and cosmology \cite{Wberg}.)

In the process of meeting the challenge, it became clear to us that
this subject would be an excellent topic for a junior or senior physics 
major's project or thesis.
After all, if a British high school student could do it\ldots
There is much more physics in the problem than just 
simply integrating a pair of coupled non-linear differential equations.
In addition to the physics (and even some astronomy), the student must 
{\it think} about the sizes of things he or she is calculating,
that is, believing and understanding the answers one gets.
Another side benefit is that the student learns about the stability of 
numerical solutions and how to deal with singularities.
In the process he or she also learns the inner 
mechanics of the calculational package (e.g., Mathematica) being used.

The student should begin with a derivation of the (Newtonian) coupled 
equations, and, presumably, be spoon-fed the general relativistic (GR) 
corrections.
Before trying to solve these equations, one needs to work out the relation 
between the energy density and pressure of the matter that constitutes
the stellar interior, i.e., an equation of state (EoS).
The first EoS's to try can be derived from the non-interacting Fermi 
gas model, which brings in quantum statistics (the Pauli exclusion principal) 
and special relativity.
It is necessary to keep careful track of dimensions, and converting to 
dimensionless quantities is helpful in working these EoS's out.

As a warm-up problem the student can, at this point, integrate the Newtonian
equations and learn about white dwarf stars.
Putting in the GR corrections, one can then proceed in the same way
to work out the structure of {\it pure} neutron stars (i.e., 
reproducing the results of Oppenheimer and Volkov \cite{TOV}).
It is interesting at this point to compare and see how important the GR 
corrections are, i.e., how different a neutron star is from that which
would be given by classical Newtonian mechanics.

Realistic neutron stars, of course, also contain some protons and 
electrons.
As a first approximation one can treat this multi-component system 
within the non-interacting Fermi gas model.
In the process one learns about chemical potentials.
To improve upon this treatment we must include nuclear interactions in
addition to the degeneracy pressure from the Pauli exclusion principle
that is used in the Fermi gas model.
The nucleon-nucleon interaction is not something we would expect
an undergraduate to tackle, but there is a simple model (which we learned 
about from Prakash \cite{Prakash}) for the nuclear matter EoS.
It has parameters which are fit to quantities such as the binding
energy per nucleon in symmetric nuclear matter, the so-called nuclear
symmetry energy (it is really an asymmetry) and the (not so well
known) nuclear compressibility.
Working this out is also an excellent exercise, which even touches on
the speed of sound (in nuclear matter).
With these nuclear interactions in addition to the Fermi gas energy
in the EoS, one finds (pure) neutron star masses and radii which are quite
different from those using the Fermi gas EoS.

The above three paragraphs provide the outline of what follows in this
paper.  In the presentation we will also indicate possible ``gotcha's'' that
the student might encounter and possible side-trips that might be taken. 
Of course, the project we outline here can (and probably should) be
augmented by the faculty mentor \cite{webpage} with suggestions for
byways that might lead to publishable results, if that is desired.

We should point out that there is a similar discussion of this subject 
matter in this journal by Balian and Blaizot \cite{BalianBlaizot}.
They, however, used this material (and other, related materials) to form the 
basis for a full-year course they taught at the Ecole Polytechnique in France.
Our emphasis is, in contrast, more toward nudging the student into a 
research frame of mind involving numerical calculation.
We also note that much of the material we discuss here is covered in 
the textbook by Shapiro and Teukolsky \cite{Shapiro}.
However, as the reader will notice, the emphasis here is on students learning 
through computation.
One of our intentions is to establish here a framework for the student
to interact with {\it his or her own} computer program, and in the
process learn about the physical scales involved in the structure of compact 
degenerate stars.

\section{The Tolman-Oppenheimer-Volkov Equation}

\subsection{Newtonian Formulation}

\begin{figure}[tbp]
        \centering
        \epsffile{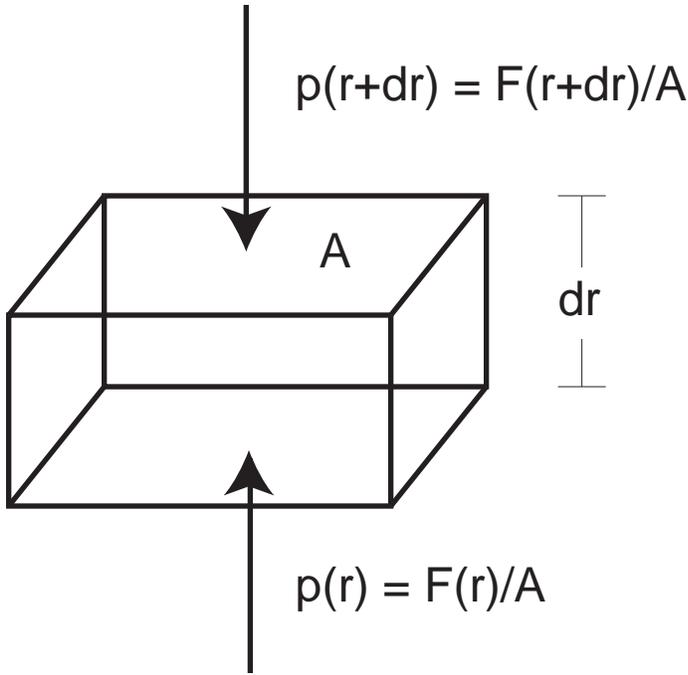}
        \caption{Diagram for derivation of Eq.\ (\ref{eq:DEpressure})}
        \label{pressureBox}
\end{figure}

A nice first exercise for the student is to derive the following structure
equations from classical Newtonian mechanics,
\begin{eqnarray} 
        \frac{d p}{d r} & = & - \frac{G \rho(r) {\cal M}(r)} {r^2} = 
        	- \frac{G \epsilon(r) {\cal M}(r)} {c^2 r^2} 
		\label{eq:DEpressure} \\ 
        \frac{d{\cal M}} {d r} & = & 4 \pi r^2 \rho(r) = 
		\frac{4 \pi r^2 \epsilon(r)} {c^2} \label{eq:DEcurlyM}\\
	{\cal M}(r)  & = &  4 \pi \int_0^r  r'^{\,2} dr' \rho(r') = 
		4 \pi \int_0^r  r'^{\,2} dr' \epsilon(r') /c^2
		\, . \label{eq:curlyM}
\end{eqnarray} 
Here $G = 6.673 \times 10^{-8}$ dyne-cm$^2$/g$^2$ is Newton's gravitational constant,
$\rho(r)$ is the mass density at the distance $r$ (in gm/cm$^3$), and 
$\epsilon$ is the corresponding energy density (in ergs/cm$^3$) \cite{CGS}.
The quantity ${\cal M}(r)$ is the total mass inside the sphere of radius
$r$.
A sufficient hint for the derivation is shown in Fig.~\ref{pressureBox}.
(Challenge question: the above equations actually hold for any value of
$r$, not just the large-$r$ situation depicted in the figure.  Can
the student also do the derivation in spherical coordinates
where the box becomes a cut-off wedge?)

Note that, in the second halves of these equations, we have departed slightly 
from Newtonian physics, defining the energy density $\epsilon(r)$ in terms 
of the mass density $\rho(r)$ according to the (almost) famous Einstein 
equation from special relativity,
\begin{equation}
	\epsilon(r) = \rho(r) c^2 \, .  \label{eq:epsdef}
\end{equation} 
This allows Eq.\ (\ref{eq:DEpressure}) to be used when one takes into account
contributions of the interaction energy between the particles making up
the star.

In what follows, we may inadvertently set $c = 1$ so that $\rho$ and 
$\epsilon$ become indistinguishable.  
We'll try not to do that here so students following the equations in 
this presentation can keeping checking dimensions as they proceed.
However, they might as well get used to this often-used physicists' trick 
of setting $c = 1$ (as well as $\hbar = 1$).

To solve this set of equations for $p(r)$ and ${\cal M}(r)$ one can integrate
outwards from the origin ($r = 0$) to the point $r = R$ where the pressure 
goes to zero.  
This point defines $R$ as the radius of the star.
One needs an initial value of the pressure at $r=0$, call it $p_0$, to do
this, and $R$ and the total mass of the star, ${\cal M}(R) \equiv M$, 
will depend on the value of $p_0$.

Of course, to be able to perform the integration, one also needs to know the
energy density $\epsilon(r)$ in terms of the pressure $p(r)$.
This relationship is the equation of state (EoS) for the matter making
up the star.
Thus, a lot of the student's effort in this project will necessarily be
directed to developing an appropriate EoS for the problem at hand.

\subsection{General Relativistic Corrections}

The Newtonian formulation presented above works well in regimes where
the mass of the star is not so large that it significantly ``warps'' 
space-time.
That is, integrating Eqs. (1) and (2) will work well in cases when
general relativistic (GR) effects are not important, such as for the 
compact stars known as white dwarfs.
By creating a quantity using $G$ that has dimensions of length, the
student can see when it becomes important to include GR effects.
(This happens when $G M/c^2 R$ becomes non-negligible.)  
As the student will see, this is the case for typical neutron stars.

It is probably not to be expected that an undergraduate physics
major derive the GR corrections to the above equations.
For that, one can look at various textbook derivations of the 
Tolman-Oppenheimer-Volkov (TOV) equation  \cite{Wberg} \cite{Shapiro}.
It should suffice to simply state the corrections to 
Eq.\ (\ref{eq:DEpressure}) in terms of three additional (dimensionless)
factors,
\begin{equation} 
        \frac{d p}{d r}  =  - \frac{G \epsilon(r) {\cal M}(r)} {c^2 r^2} 
		\left[ 1 + \frac{p(r)}{\epsilon(r)} \right]
		\left[1 + \frac{4 \pi r^3 p(r)} {{\cal M}(r) c^2} \right]
		\left[1 - \frac{2 G {\cal M}(r)} {c^2 r} \right]^{-1} \ .
		\label{eq:DEpressureGR}
\end{equation}
The differential equation for ${\cal M}(r)$ remains unchanged.
The first two factors in square brackets represent special relativity
corrections of order $v^2/c^2$.  
This can be seen in that pressure $p$ goes, in the non-relativistic
limit, like $k_F^2/2m = mv^2/2$ (see Eq.\ ({\ref{eq:electronpres}) below)
while $\epsilon$ and ${\cal M}c^2$ go like $mc^2$.
That is, these factors reduce to 1 in the non-relativistic limit.
(The student should have, by now, realized that $p$ and $\epsilon$ have the
same dimensions.)
The last factor is a GR correction and the size of $G{\cal M}/c^2r$,
as we emphasized above, determines whether it is important (or not).

Note that the correction factors are all positive definite.
It is as if Newtonian gravity becomes stronger at every $r$.
That is, special and general relativity strengthens the relentless 
pull of gravity!

The coupled non-linear equations for $p(r)$ and ${\cal M}(r)$ can
also in this case be integrated from $r=0$ for a starting value of 
$p_0$ to the point where $p(R)=0$, to determine the star mass
$M={\cal M}(R)$ and radius $R$ for this value of $p_0$.  
These equations invoke a balance between gravitational 
forces and the internal pressure.
The pressure is a function of the EoS, and for certain conditions
it may not be sufficient to withstand the gravitational attraction.
Thus the structure equations imply there is a maximum mass that a
star can have.

\section{White Dwarf Stars}

\subsection{A Few Facts}

Let us violate (in words only) the second law of
thermodynamics by warming up on cold compact stars called white dwarfs.
For these stellar objects, it suffices to solve the Newtonian structure 
equations, Eqs.\ (\ref{eq:DEpressure})-(\ref{eq:curlyM}) \cite{Koonin}.

White dwarf stars \cite{NASA} were first observed in 1844 by Friedrich 
Bessel (the same person who invented the special functions bearing 
that name).
He noticed that the bright star Sirius wobbled back and forth and
then deduced that the visible star was being orbited by some unseen
object, i.e., it is a binary system.
The object itself was resolved optically some 20 years later and
thus {\it earned} the name of ``white dwarf.''
Since then numerous other white (and the smaller brown) dwarf stars have been
observed (or detected).

A white dwarf star is a low- or medium-mass star near the end of its 
lifetime, having burned up, through nuclear processes, most of its hydrogen 
and helium forming carbon, silicon or (perhaps) iron.  
They typically have a mass less than 1.4 times that of our Sun, 
${\rm M}_\odot = 1.989 \times 10^{33}$ g \cite{Chandra}. 
They are also much smaller than our Sun, 
with radii of the order of $10^4$ km (to be compared with 
$R_\odot = 6.96 \times 10^5$ km).
These values can be worked out from the period of the wobble for
the dwarf-normal star binary in the usual Keplerian way.
As a result (and as is also the case for neutron stars), the natural
dimensions for discussing white dwarfs are for masses to be in 
units of solar mass, ${\rm M}_\odot$, and distances to be in km.
Using these numbers the student should be able to make a quick estimate 
of the (average) densities of our Sun and of a white dwarf, to get a
feel for the numbers that he will be encountering.

Since $GM/c^2R \approx 10^{-4}$ for such a typical white dwarf, 
we can concentrate here on solving the
non-relativistic structure equations of Sec. 2.1.
(Question: why is it a good approximation to drop the special
relativistic corrections for these dwarfs?)

The reason a dwarf star is small is because, having burned up all the
nuclear fuel it can, there is no longer enough thermal pressure to 
prevent its gravity from crushing it down.
As the density increases, the electrons in the atoms are pushed
closer together, which then tend to fall into the lowest
energy levels available to them. 
(The star begins to get colder.)
Eventually the Pauli principle takes over, and the electron degeneracy
pressure (to be discussed next) provides the means for stabilizing
the star against its gravitational attraction \cite{Chandra,Shapiro}.
This is the physics behind the EoS which one needs to integrate the
Newtonian structure equations above, 
Eqs.\ (\ref{eq:DEpressure}) and (\ref{eq:DEcurlyM}).

\subsection{Fermi Gas Model for Electrons}

For free electrons the number of states $d n$ available at momentum
$k$ per unit volume is
\begin{equation}
	d n = \frac{d^3 k}{(2 \pi \hbar)^3} =
		\frac{ 4 \pi k^2 d k}{(2 \pi \hbar)^3} \, .
\end{equation}
(This is a result from their modern physics course that  students
should review if they don't remember it.)  
Integrating, one gets the electron number density
\begin{equation}
	n = \frac{ 8 \pi}{(2 \pi \hbar)^3} \int_0^{k_F}  k^2 d k 
		= \frac{k_F^3}{3 \pi^2 \hbar^3}   \, . \label{eq:nofkF}
\end{equation}
The additional factor of two comes in because there are two spin states 
for each electron energy level. 
Here $k_F c$, the Fermi energy, is the maximum energy electrons can have
in the star under consideration.
It is a parameter which varies according to the star's total mass
and its history, but which the student is free to set in the
calculations he or she is about to make.

Each electron is neutralized by a proton, which in turn is accompanied
in its atomic nucleus by a neutron (or perhaps a few more, as in the case
of a nucleus like $^{56}{\rm Fe}_{26}$).
Thus, neglecting the electron mass $m_e$ with respect to the nucleon
mass $m_N$, the mass density of the star is essentially given by
\begin{equation}
	\rho = n m_N A/Z \, ,
\end{equation}
where $A/Z$ is the number of nucleons per electron.
For $^{12}{\rm C}$, $A/Z = 2.00$, while for $^{56}{\rm Fe}$, $A/Z = 2.15$ .
Note that, since $n$ is a function of $k_F$, so is $\rho$.
Conversely, given a value of $\rho$,
\begin{equation}
	k_F = \hbar\left(\frac{3\pi^2\rho}{m_N}\frac{Z}{A}\right)^{1/3} \, .
\end{equation}
The energy density of this star is also dominated by the nucleon masses,
i.e., $\epsilon \approx \rho c^2$.
 
The contribution to the energy density from the electrons (including their 
rest masses) is 
\begin{eqnarray}
	\epsilon_{\rm elec}(k_F) & = & \frac{8 \pi}{(2 \pi \hbar)^3} 
		\int_0^{k_F} (k^2 c^2 + m_e^2 c^4)^{1/2} k^2 d k 
		\nonumber \\
		& = & \epsilon_0 \int_0^{k_F/m_e c} (u^2 + 1)^{1/2} u^2 d u 
		\nonumber \\
		& = & \frac{\epsilon_0}{8} \left[ (2 x^3 + x)(1 + x^2)^{1/2}
			- \sinh^{-1}(x) \right] 
		\label{eq:electroneps}	\, ,
\end{eqnarray}
where
\begin{equation}
	\epsilon_0 = \frac{m_e^4 c^5}{\pi^2 \hbar^3}
\end{equation}
carries the desired dimensions of energy per volume and $x = k_F / m_e c$.
The {\it total} energy density is then
\begin{equation}
	\epsilon = n m_N A/Z + \epsilon_{\rm elec}(k_F)  \, .
	\label{eq:totaleps}
\end{equation}
One should check that the first term here is much larger than the second.

To get our desired EoS, we need an expression for the pressure.
The following presents a problem (!) that the student should work through.
From the first law of thermodynamics, $dU = dQ - pdV$,
then at temperature $T$ fixed at $T=0$ 
(where $dQ = 0$ since $dT = 0$)
\begin{equation}
	p = - \left. \frac{\partial U}{\partial V} \right]_{T=0} =
		n^2 \frac{d(\epsilon/n)} {d n} = 
		n \frac{d \epsilon}{d n} - \epsilon =
		n \mu - \epsilon  \, ,  \label{eq:pressureDef}
\end{equation}
where the energy density here is the total one given by 
Eq.\ (\ref{eq:totaleps}).
The quantity $\mu_i = d\epsilon/dn_i$ defined in the last equality is known 
as the  chemical potential of the electrons.
This is a concept which will be especially useful in  
Section 5 where we consider an equilibrium mix of neutrons, protons 
and electrons.

Utilizing Eq.\ (\ref{eq:electroneps}), Eq.\ (\ref{eq:pressureDef}) yields
the pressure (another problem!)
\begin{eqnarray}
	 p(k_F) & = & \frac{8 \pi}{3 (2 \pi \hbar)^3} 
		\int_0^{k_F} (k^2 c^2 + m_e^2 c^4)^{-1/2} k^4 d k 
		\nonumber \\
		& = & \frac{\epsilon_0}{3} \int_0^{k_F/m_e c} 
			(u^2 + 1)^{-1/2} u^4 d u 
		\nonumber \\
		& = & \frac{\epsilon_0}{24} \left[ (2 x^3 -3 x)(1 + x^2)^{1/2}
			+3 \sinh^{-1}(x) \right]
		\label{eq:electronpres}	\, .
\end{eqnarray}
(Hint: use the $n^2 d(\epsilon/n)/dn$ form and remember to integrate
by parts.)

Using Mathematica \cite{Mma} the student can show that the constant 
$\epsilon_0 = 1.42 \times 10^{24}$ in units that, at this point,
are erg/cm$^3$.
(Yet another problem: verify that the units of 
$\epsilon_0$ are as claimed \cite{workouteqns}.)
One also finds that Mathematica can perform the integrals analytically.
(We quoted the results already in the equations above.)
They are a bit messy, however, as they both involve an inverse hyperbolic sine
function, and thus are not terribly enlightening.  
It is useful, however, for the student to make a plot of $\epsilon$ versus 
$p$ (such as shown in Fig.\ \ref{electronPVsEps}) for 
values of the parameter $0 \leq k_F \leq 2 m_e$.
\begin{figure}[tbp]
        \centering
	$p(\epsilon)$\\
        \epsffile{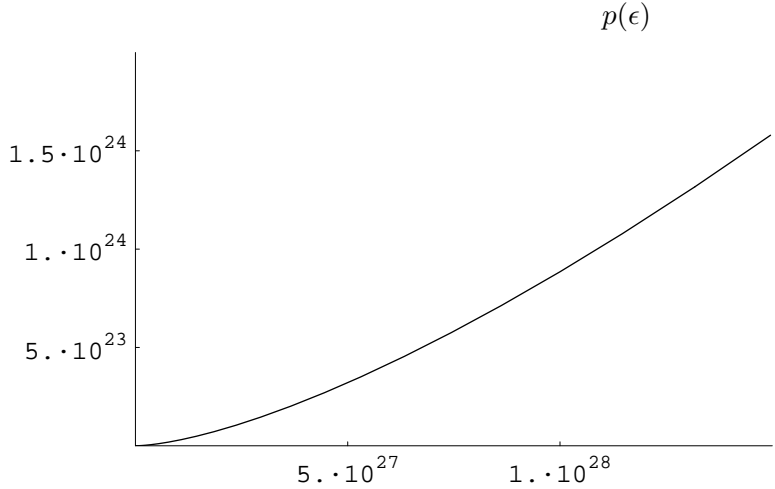}
        \caption{Relation between pressure $p$ ($y$-axis) and
		energy density $\epsilon$ ($x$-axis)
		in the free electron Fermi gas model.
		Units are ergs/cm$^3$.  Note that the pressure is much
		smaller than the energy density, since the latter is
		dominated by the massive nucleons.}
        \label{electronPVsEps}
\end{figure} 
This curve has a shape much like $\epsilon^{4/3}$ (the student should
compare with this), and there is a good reason for that.

Consider the (relativistic) case when $k_F \gg m_e$.  Then 
Eq.\ (\ref{eq:electronpres}) simplifies to
\begin{eqnarray}
	p(k_F) & = & \frac{\epsilon_0}{3} \int_0^{k_F/m_e c} u^3 d u =
		\frac{\epsilon_0}{12}(k_F/m_e c)^4 =
		\frac{\hbar c}{12 \pi^2} 
			\left( \frac{3 \pi^2 Z \rho}{m_N A} \right)^{4/3}
			\nonumber \\
		& \approx & K_{\rm rel} \,  \epsilon^{4/3} \, ,
	\label{eq:elpresRelvstc}
\end{eqnarray}
where
\begin{equation}
	K_{\rm rel} = \frac{\hbar c}{12 \pi^2}
		\left( \frac{3 \pi^2 Z}{A m_N c^2} \right)^{4/3} \, .
	\label{eq:Krelvstc}
\end{equation}
A star having simple EoS like $p = K \epsilon^\gamma$ is called a 
``polytrope'', and we therefore see that the relativistic electron Fermi 
gas gives a polytropic EoS with $\gamma = 4/3$.
As will be seen in the next subsection, a polytropic EoS allows one to 
solve the structure equations (numerically) in a relatively 
straight-forward way \cite{LaneEmden}.

There is another polytropic EoS for the non-interacting electron Fermi 
gas model corresponding to the non-relativistic limit, where $k_F \ll m_e$.  
In a way similar to the derivation of Eq.\ (\ref{eq:elpresRelvstc}), one finds
\begin{equation}
	p = K_{\rm non-rel} \epsilon^{5/3} \, , \quad {\rm where} \ \ 
	K_{\rm non-rel} = \frac{\hbar^2}{15 \pi^2 m_e}
		\left( \frac{3 \pi^2 Z}{A m_N c^2} \right)^{5/3} \, .
	\label{eq:Knonrelvstc}
\end{equation}
[Question: what are the units of $K_{\rm rel}$ and $K_{\rm non-rel}$?
Task: confirm that in the appropriate limits, Eqs.\ (\ref{eq:electroneps}) 
and (\ref{eq:electronpres}) reduce to those in Eqs.\ (\ref{eq:elpresRelvstc}) 
and (\ref{eq:Knonrelvstc}).]

\subsection{The Structure Equations for a Polytrope}

As mentioned earlier, we want to express our results in units of km
and ${\rm M}_\odot$.  Thus it is useful to define $ \bar{\cal M}(r) =
{\cal M}(r) / {\rm M}_{\odot} $.  The first Newtonian structure
equation, Eq.\ (\ref{eq:DEpressure}), then becomes
\begin{equation}
	\frac{d p(r)}{d r} = -R_0 \frac{\epsilon(r) \bar{\cal M}(r)}{r^2} \, ,
	\label{eq:DEpresDimlessM}
\end{equation}
where the constant $R_0 = G {\rm M}_\odot/c^2 = 1.47$ km.
(That is, for those who already know, $R_0$ is one half the 
Schwartzschild radius of our sun.)
In this equation $p$ and $\epsilon$ still carry dimensions of, say,
ergs/cm$^3$.
Therefore, let us define dimensionless energy density and pressure,
$\bar\epsilon$ and $\bar{p}$, by
\begin{equation}
	p = \epsilon_0 \bar{p} \, , \quad 
	\epsilon = \epsilon_0 \bar{\epsilon} \,
	\label{eq:epspDimless}
\end{equation}
where $\epsilon_0$ has dimensions of energy density. 
This $\epsilon_0$ is not the same as defined in Eq.\ (11).
Its numerical choice here is arbitrary, and a suitable strategy is to make that
choice based on the dimensionful numbers that define the problem at hand.
We'll employ this strategy to fix it below.  For a polytrope, we can write
\begin{equation}
	\bar{p} = \bar{K} \bar{\epsilon}^{\,\gamma} \, , \quad {\rm where} \ \ 
	\bar{K} = K \epsilon_0^{\ \gamma - 1} \, {\rm\ is\  dimensionless.}
	\label{eq:polytropeDimless}
\end{equation}

It is easier to solve Eq.\ (\ref{eq:DEpresDimlessM}) for $\bar{p}$, so
we should express $\bar{\epsilon}$ in terms of it,
\begin{equation}
	\bar{\epsilon} = (\bar{p}/\bar{K})^{1/\gamma} \, .
	\label{eq:epsbarofp}
\end{equation}
Equation (\ref{eq:DEpresDimlessM}) can now be recast in the form
\begin{equation}
	\frac{d \bar{p}(r)}{d r} = 
		- \frac{\alpha\, \bar{p}(r)^{1/\gamma} \bar{\cal M}(r)}
			{r^2} \, , \label{eq:DEpresDimless}
\end{equation}
where the constant $\alpha$ is
\begin{equation}
	\alpha = R_0/\bar{K}^{1/\gamma} = 
		R_0/(K \epsilon_0^{\gamma - 1})^{1/\gamma} \, .
	\label{eq:alphadef}
\end{equation}
Equation (\ref{eq:DEpresDimless}) has dimensions of 1/km, with
$\alpha$ in km (since $R_0$ is).
That is, it is to be integrated with respect to $r$, with $r$ also in km.

We can choose any convenient value for $\alpha$ since $\epsilon_0$ is still
free.
For a given value of $\alpha$, $\epsilon_0$ is then fixed at
\begin{equation}
	\epsilon_0 = \left[ \frac{1}{K} \left( \frac{R_0}{\alpha}
		\right)^\gamma \right]^{1 - \gamma} \, .
	\label{eq:eps0def}
\end{equation}

We also need to cast the other coupled equation, Eq.\ (\ref{eq:DEcurlyM}),
in terms of dimensionless $\bar{p}$ and $\bar{\cal M}$,
\begin{equation}
	\frac{d \bar{\cal M}(r)}{dr} = \beta r^2 \bar{p}(r)^{1/\gamma} \, ,
	\label{eq:DEcurlyMDimless}
\end{equation}
where \cite{notCoul}
\begin{equation}
	\beta = \frac{4\pi\epsilon_0}{{\rm M}_\odot c^2 \bar{K}^{1/\gamma}} =
		\frac{4\pi\epsilon_0}
			{{\rm M}_\odot c^2 (K \epsilon_0^{\gamma-1})^{1/\gamma}}
		 \, . \label{eq:betadef}
\end{equation}
Equation (\ref{eq:DEcurlyMDimless}) also carries dimensions of 1/km,
the constant $\beta$ having dimesnions 1/km$^3$.
Note that, in integrating out from $r=0$, the initial value of
$\bar{\cal M}(0) = 0$.

\subsection{Integrating the Polytrope Numerically}

Our task is to integrate the coupled first-order differential equations
(DE), Eqs.\ (\ref{eq:DEpresDimless}) and (\ref{eq:DEcurlyMDimless}), 
out from the origin, $r=0$, to the point $R$ where the pressure falls
to zero, $\bar{p}(R) = 0$ \cite{monofall}.
To do this we need two initial values, $\bar p(0)$ (which must be 
positive) and $\bar{\cal M}(0)$ ( which we already know must be 0).
The star's radius, $R$, and its mass $M = \bar{\cal M}(R)$ in units
of ${\rm M}_\odot$ will vary, depending on the choice for $\bar p(0)$.

For purposes of numerical stability in solving Eqs.\ (\ref{eq:DEpresDimless}) 
and (\ref{eq:DEcurlyMDimless}), we want the constants $\alpha$ and
$\beta$ to be not much different from each other (and probably not much
different from 1).
We will see below that this can be arranged for both of the two
polytropic EoS's discussed above for white dwarfs.

Our coupled DE's are quite non-linear.
In fact, because of the $\bar{p}^{1/\gamma}$ factors, the solution will
become complex when $\bar{p}(r) < 0$, i.e., when $r > R$.
Thus we will want to recognize when this happens.
How can this be programmed?

Mathematica and similar symbolic/numerical packages have built-in
first-order DE solvers.
Perhaps the solver is as simple as a fixed, equal-step Runge-Kutta
routine (as in MathCad 7 Standard), but there are often more sophisticated
solvers in more recent versions.
These packages also allow for program control constructs such as 
do-loops, whiles and the like.

Thus, consider a do-loop on a variable $\bar{r}$ running in appropriately
small steps over a range that is sure to contain the expected value of $R$.
Call the DE solver inside this loop, integrating the coupled DE's from
$r=0$ to $\bar{r}$.
When the solver routine exits, check to see if the last value of $\bar{p}$,
i.e., $\bar{p}(\bar{r})$, has a real part which has gone negative.
If so, then break out of the loop, calling $R = \bar{r}$.
If not, go on to the next larger value of $\bar{r}$ and call the
DE solver again.

More discussion of how to program the integration of the DE's is
inappropriate here, since we want to encourage the student to learn from
programming to appreciate how the symbolic/numerical package is used.

\subsection{The Relativistic Case $k_F \gg m_e$}

This is the regime for white dwarfs with the largest mass.
A larger mass needs a greater central pressure to support it.
However, large central pressures mean the squeezed electrons become
relativistic.

Recall that the polytrope exponent $\gamma = 4/3$ for this case and
the equation of state is given by $ P = K_{\rm rel} \epsilon^{\gamma}$
with $K_{\rm rel}$ given by Eq.\ (\ref{eq:Krelvstc}).  After some
trial and error, we chose in our program (the student may want to try
something else)
\begin{equation}
	\alpha = R_0 = 1.473 \ {\rm km} \quad\quad\quad [k_F \gg m_e] \, ,
\end{equation}
which in turn fixes, from Eq.\ (\ref{eq:eps0def}),
\begin{equation}
	\epsilon_0 = 7.463 \times 10^{39} {\rm ergs/cm}^3 =
		4.17 \ {\rm M}_\odot c^2/{\rm km}^3
		\quad\quad\quad [k_F \gg m_e] \, .
\end{equation}
The first question the student should ask, in checking this number, 
is whether such a large number is physically reasonable.

Continuing with the $k_F \gg m_e$ numerics, Eqs.\ (\ref{eq:Krelvstc}) 
and (\ref{eq:betadef}) give
\begin{equation}
	\beta = 52.46 \, /{\rm km}^3  \quad\quad\quad [k_F \gg m_e] \, ,
\end{equation}
which is about 30 times larger than $\alpha$, but probably manageable
from the standpoint of performing the numerical integration.

In {\it our} first attempt to integrate the coupled DE's for this case
(using a do-loop as described above) we chose $\bar{p}(0) = 1.0$.
This gives us a white dwarf of radius $R \approx 2$ km, which is
miniscule compared with the expected radius of $\approx 10^4$ km!
Why?  What went wrong?

The student who also makes this kind of mistake will eventually
realize that our choice of scale, 
$\epsilon_0 = 4.17 {\rm M}_\odot c^2/{\rm km}^3$,
represents a {\it huge} energy density.
One can simply estimate the average energy density of a star with 
a $10^4$ km radius and a mass of one solar mass by the ratio of its rest 
mass energy to its volume,
\begin{equation}
	\left<\epsilon\right> \approx \frac{{\rm M}_\odot c^2}{R^3} = 
		10^{-12}  \ {\rm M}_\odot c^2/{\rm km}^3   \, ,
\end{equation}
which is much, much smaller than the $\epsilon_0$ here.
In addition, the pressure $p$ is about 2000 times smaller than the
energy density $\epsilon$ (see Fig.\ \ref{electronPVsEps}).
Thus, choosing a starting value of $\bar{p}(0) \sim 10^{-15}$ would probably
be more physical.

Doing so does give much more reasonable results.
Table \ref{tab:WDrel} shows our program's results for $R$ and $M$ and 
how they depend on $\bar{p}(0)$.
\begin{table}
\centering
\caption{Radius $R$ (in km) and mass $M$ (in ${\rm M}_\odot$) for white 
	dwarfs with a relativistic electron Fermi gas EoS.}

\label{tab:WDrel}
\begin{tabular}{|r|r|r|}
\hline
$\bar{p}(0) \ $ 	& $R$  	& $M$ \\
\hline
$10^{-14}$	& 4840		& 1.2431 \\
\hline
$10^{-15}$	& 8600		& 1.2432 \\
\hline
$10^{-16}$	& 15080		& 1.2430 \\
\hline
\end{tabular}
\end{table}
The surprise here is that, within the numerical error expected, 
all these cases have the {\it same} mass!
Increasing the central pressure doesn't allow the star to be more
massive, just more compact.

It turns out that this result is correct, i.e., that the white dwarf
mass is independent of the choice of the central pressure.  It is not
easy to see this, however, from the numerical integration we have done
here.  The discussion in terms of Lane-Emden functions \cite{LaneEmden} 
shows why, though the mathematics here might be a bit
steep for many undergraduates. For this reason, we quote without
proof the analytic results. For the case
of a polytropic equation of state $p=K \epsilon^\gamma$ , the mass
\begin{equation}
M = 4 \pi \epsilon^{2(\gamma-\frac{4}{3})/3}~\left(\frac{K \gamma}{4 \pi G(\gamma-1)}\right)^{3/2}~\zeta_1^2~ |\theta(\zeta_1)| \,,
\label{eq:laneemdenmass}
\end{equation}
and the radius
\begin{equation}
R = \sqrt{\frac{K \gamma}{4 \pi G(\gamma-1)}}~\zeta_1~\epsilon^{(\gamma-2)/2} 
\,.
\label{eq:laneemdenradius}
\end{equation} 
In the above-mentioned solutions, $\zeta_1$ and $\theta(\zeta_1)$ are
numerical constants that depend on the polytropic index $\gamma$.  By
examining Eq. \ (\ref{eq:laneemdenmass}), we see that for $\gamma = 4/3$
the mass is independent of the central energy density, and hence also
the central pressure $p_0$.  Also,
note that from Eq. \ (\ref{eq:laneemdenradius}), the radius decreases
with increasing central pressure as $ R \propto
p_0^{(\gamma-2)/2\gamma} = p_0^{-1/4}$.  In any case, the student should notice
this point and use it as check of the numerical results obtained. 
Figure \ref{WDrelGraphs} shows the dependence of $\bar{p}(r)$ and 
$\bar{\cal M}(r)$ on distance for the case $\bar{p}(0) = 10^{-16}$.
\begin{figure}[tbp]
        \centering
        \epsffile{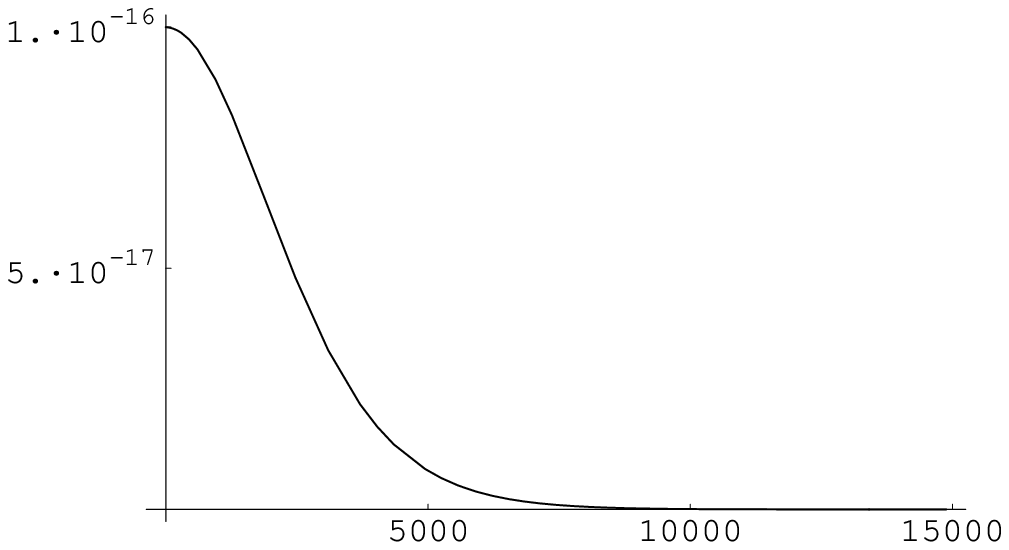}
        \epsffile{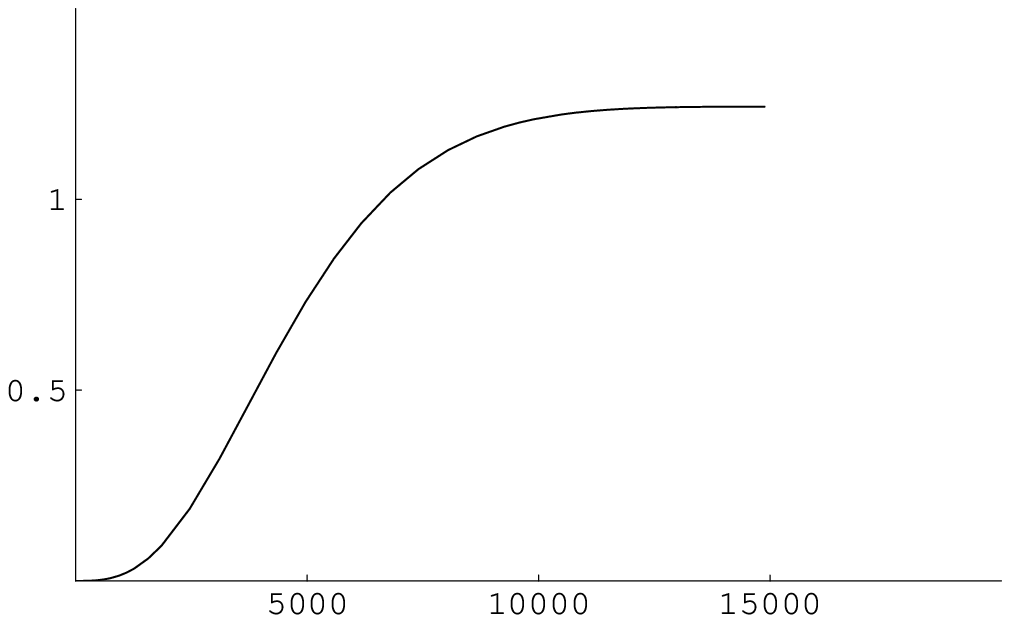}
        \caption{ $\bar{p}(r)$ and $\bar{\cal M}(r)$ for 
	white dwarfs using the relativistic 
	electron Fermi gas model.  Here $\bar{p}(0) = 10^{-16}$.}
        \label{WDrelGraphs}
\end{figure} 
It is interesting that $\bar{p}(r)$ becomes small and essentially
flat around 8000 km before finally going through zero at $R = 15,080$ km.

The results and graphs shown here were generated with a Mathematica 4.0
program, but we were able to reproduce them using MathCad 7 Standard.
In that case, however, programming a loop is difficult, so we searched by hand
for the endpoint (where the real part of $\bar{p}(r)$ goes negative).
More recent versions of MathCad have more complete program constructs, 
such as while-loops, so this process could undoubtedly be automated.
(Alternatively, the student might try to solve for a root of 
${\bar p}(r) = 0$.)

\subsection{The Non-Relativistic Case, $k_F \ll m_e$}

Eventually, as the central pressure $\bar{p}(0)$ gets smaller, the
electron gas is no longer relativistic.
Also as the pressure gets smaller, it can support less mass.
This moves us in the direction of the less massive white dwarfs,
and, as it turns out, these dwarfs are {\it larger} (in radius) than
the ones in the last section.

In the extreme case, when  $k_F \ll m_e$, we can integrate the structure
equations for the other polytropic EoS, where $\gamma = 5/3$.
The programming for this is very much the same as in the 4/3 case,
but the numbers involved are quite different (as are the results).

Inserting the values of the physical constants in Eq.\ (\ref{eq:Knonrelvstc}),
we find
\begin{equation}
	K_{\rm non-rel} = 3.309 \times 10^{-23} \
		\frac{{\rm cm}^2}{{\rm ergs}^{2/3}} \, .
\end{equation}
This time, however, and after some experimentation, we chose the constant
\begin{equation}
	\alpha = 0.05 \ {\rm km} \quad\quad\quad [k_F \ll m_e] \, ,
\end{equation}
which then fixes
\begin{equation}
	\epsilon_0 = 2.488 \times 10^{37} {\rm ergs/cm}^3 =
		0.01392 \ {\rm M}_\odot c^2/{\rm km}^3
		\quad\quad\quad [k_F \ll m_e] \, .
\end{equation}
Note that this $\epsilon_0$ is much smaller than our choice
for the relativistic case.
The other constant we need, from Eq.\ (\ref{eq:betadef}), is
\begin{equation}
	\beta = 0.005924 \ /{\rm km}^3  \quad\quad\quad [k_F \ll m_e] \, ,
\end{equation}
which, unlike the relativistic case, is not larger than $\alpha$ but smaller.

When we first ran our Mathematica code for this case, we (inadvertantly)
tried a value of $\bar{p}(0) = 10^{-12}$.  
This gave us a star with radius $R$ = 5310 km and mass $M$ = 3.131.
Oops!, that mass is {\it bigger} than the largest mass of 1.243 that we
found for the relativistic EoS!
What did we do wrong?

What happened (and the student can set up her program so this trap
can be avoided) is that the choice $\bar{p}(0) = 10^{-12}$ violates
the assumption that $k_F \ll m_e$.
One really needs values for $\bar{p}(0) < 4 \times 10^{-15}$.
This says, in fact, that the relativistic $\bar{p}(0) = 10^{-16}$ 
case that we plotted in Fig.\ \ref{WDrelGraphs} is not really relativistic.

Results for the non-relativistic case for the last two values of
$\bar{p}(0)$ in Table \ref{tab:WDrel} are shown in Table \ref{tab:WDnonrel}.
\begin{table}
\centering
\caption{Radius $R$ (in km) and mass $M$ (in ${\rm M}_\odot$) for white 
	dwarfs with a non-relativistic electron Fermi gas EoS.}

\label{tab:WDnonrel}
\begin{tabular}{|r|r|r|}
\hline
$\bar{p}(0) \ $ & $R$ 	& $M$ \\
\hline
$10^{-15}$	& 10620		& 0.3941 \\
\hline
$10^{-16}$	& 13360		& 0.1974 \\
\hline
\end{tabular}
\end{table}
It is quite instructive to compare the differences in the two tables.
The masses are, of course, smaller, as expected, and now they vary
with $\bar{p}(0)$.  
Somewhat surprising is that the non-relativistic radius is bigger for 
$\bar{p}(0) = 10^{-15}$  but smaller for  $\bar{p}(0) = 10^{-16}$.
Figure \ref{WDnonrelGraphs} shows the pressure distribution for the latter 
case, to be compared with the corresponding graph in Fig. \ref{WDrelGraphs}.
\begin{figure}[tbp]
        \centering
	$\bar{p}(r)$\\
        \epsffile{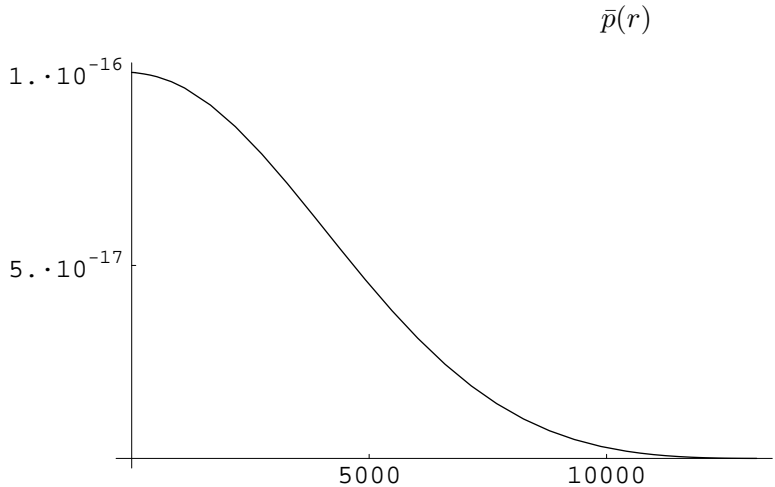}
        \caption{ $\bar{p}(r)$ for a white dwarf using the non-relativistic 
	electron Fermi gas model with central pressure 
	$\bar{p}(0) = 10^{-16}$.}
        \label{WDnonrelGraphs}
\end{figure} 
Note that this pressure curve does not have the peculiar, long flat tail
found using the relativistic EoS.

In fact, by this time the student should have realized that neither
of these polytropes is very physical, at least not for all cases.
The non-relativistic assumption certainly does not work for central pressures
$\bar{p}(0) > 10^{-14}$, i.e., for the more massive (and more common)
white dwarfs.
On the other hand, the relativistic EoS certainly should not work
when the pressure becomes small, i.e., in the outer regions of the
star (where it eventually goes to zero at the star's radius).
So, can one find an EoS to cover the whole range of pressures?

We haven't actually done this for white dwarfs, but the program would be 
much like that discussed below for the full neutron star.
Given the transcendental expressions for energy and pressure that 
generate the curve shown in Fig. \ref{electronPVsEps}, it should be possible to
find a fit (using, for example, the built-in fitting function of 
Mathematica) like
\begin{equation}
	\bar\epsilon(\bar p) = A_{\rm NR}\bar{p}^{3/5} + A_{\rm R}\bar{p}^{3/4}
		\, .  \label{eq:WDfitEoS}
\end{equation}
The second term dominates at high pressures (the relativistic case),
but the first term takes over for low pressures when the $k_F \gg m_e$
assumption does not hold.
(Setting the two terms equal and solving for $\bar p$, as Chandrasekhar and
Fowler did, gives the value of $\bar p$ when special relativity starts
to be important.)
This expression for $\bar\epsilon(\bar p)$ could then be used in place of 
the $\bar{p}^{1/\gamma}$
factors on the right hand sides of the structure equations.
Proceed to solve numerically as before.
We leave this as an exercise for the interested student.

\section{Pure Neutron Star, Fermi Gas EoS}

Having by now become warm, the student can now tackle neutron stars.
Here one must include the general relativistic (GR) contributions
represented by the three dimensionless factors in the TOV equation,
Eq.\ (\ref{eq:DEpressureGR}).
One of the first things that comes to mind is how one deals numerically 
with the (apparent) singularities in these factors at $r=0$ \cite{hint}.

Also, as in the case of the white dwarfs, there is a question of what 
to use for the EoS.
In this section we show what can be done for {\it pure} neutron stars, 
once again using a Fermi gas model for, now, a neutron gas instead of
an electron gas.
Such a model, however, is unrealistic for two reasons.
First, a real neutron star consists not just of neutrons but contains
a small fraction of protons and electrons (to inhibit the neutrons
from decaying into protons and electrons by their weak interactions).
Second, the Fermi gas model ignores the strong nucleon-nucleon interactions, 
which give important contributions to the energy density.
Each of these points will be dealt with in sections below.

\subsection{The Non-Relativistic Case, $k_F \ll m_n$}

For a pure neutron star Fermi gas EoS one can proceed much as in the white 
dwarf case, substituting the neutron mass $m_n$ for the electron mass $m_e$ 
in the equations found in Sec.\ 3.
When $k_F \ll m_n$ one finds, again, a polytrope  with $\gamma = 5/3$.
(More exercises for the student.)
The $K$ corresponding to that in Eq.\ (\ref{eq:Knonrelvstc}) is 
\begin{equation}
	K_{\rm non-rel} = \frac{\hbar^2}{15 \pi^2 m_n}
		\left( \frac{3 \pi^2 Z}{A m_n c^2} \right)^{5/3} =
		6.483 \times 10^{-26} \ \frac{{\rm cm}^2}{{\rm ergs}^{2/3}}
		  \, . \label{eq:KnonrelNstar}
\end{equation}
This time, choosing $\alpha = 1$ km, one finds the scaling factor from 
Eq.\ (\ref{eq:eps0def}) to be
\begin{equation}
	\epsilon_0 = 1.603 \times 10^{38} \ {\rm ergs/cm}^3 =
		0.08969 \  {\rm M}_\odot c^2/{\rm km}^3 \, .
	\label{eq:eps0nonrelNstar}
\end{equation}
Further, from Eqs.\ (\ref{eq:polytropeDimless}) and (\ref{eq:betadef}),
\begin{equation}
	\bar{K} = 1.914 \quad {\rm and} \ 
	\beta = 0.7636 \ /{\rm km}^3 \, . \label{eq:KbarBetanonrelNstar}
\end{equation}
Note that, in this case, the constants $\alpha$ and $\beta$ are of 
similar size.

Making an estimate of the average energy density of a typical neutron
star (mass = ${\rm M}_\odot$, $R$ = 10 km), one expects that a good 
starting value for the central pressure $\bar{p}(0)$ to be of order
$10^{-4}$ or less.
Our program for this situation is essentially the same as the one for
non-relativistic white dwarfs but with appropriate changes of the
distance scale.
It gives the results shown in Table \ref{tab:nonrelNstar}.
\begin{table}
\centering
\caption{Radius $R$ (in km) and mass $M$ (in M$_\odot$) for pure neutron stars 
	with a non-relativistic Fermi gas EoS.}

\label{tab:nonrelNstar}
\begin{tabular}{|c|c|c|c|c|}
\hline
$\bar{p}(0) \ $ & $R$ (Newton) & $M$ (Newton) 
		& $R$ (GR) & $M$ (GR)\\
\hline
$10^{-4}$	& 16.5 & 0.7747 & 15.25 & 0.6026 \\
\hline
$10^{-5}$	& 20.8 & 0.3881 & 20.00	& 0.3495 \\
\hline
$10^{-6}$	& 26.3 & 0.1944 & 25.75 & 0.1864 \\
\hline
\end{tabular}
\end{table}
Note that the GR effects are small, but not negligible, for this 
non-relativistic EoS.
As in the white dwarf case, these are the smaller mass stars.
One sees that as the mass gets smaller, the gravitational attraction is 
less and thus the star extends out to larger radii.

\subsection{The Relativistic Case, $k_F \gg m_n$}

Here there is again a polytropic EoS, but with $\gamma = 1$.
In fact, $p = \epsilon/3$, a well-known result for a relativistic gas.
The conversion to dimensionless quantities becomes very simple in this case
with relationships like $K = \bar{K} = 1/3$.
It is still useful to factor out an $\epsilon_0$, which in our
program we took to have a value $1.6\times 10^{38}$ erg/cm$^3$, as suggested
by the value in the previous sub-section.
Then, if we choose this time
\begin{equation}
	\alpha = 3 R_0 = 4.428 \ {\rm km}
\end{equation}
we find 
\begin{equation}
	\beta = 3.374 \ /{\rm km}^3  \, .
\end{equation}
We expect central pressures $\bar{p}(0)$ in this case to be greater
than $10^{-4}$.
Other than these changes, we wrote a similar program to the one above,
taking care to avoid exponents like $1/(\gamma - 1)$.

Running that code gives, at first glance, {\it enormous} 
radii, values of $R$ greater than 50 km!
We can imagine the student looking frantically for a program
bug that isn't there.
In fact, what really happens is that, for this EoS, the loop on
$\bar r$ runs through its whole range, since the pressure $\bar{p}(r)$
never passes through zero.
(A plot of $\bar{p}(r)$ looks quite similar, but for distance scale, 
to that shown in Fig.\ \ref{WDrelGraphs}, where $\gamma = 4/5$.)
It only falls monotonically toward zero, getting ever smaller.
By the time the student recognizes this, she will probably also have
realized that the relativistic gas EoS is inappropriate for such
small pressures.
Something better should be done (as in the next sub-section).

It turns out that the structure equations for $\gamma = 1$ are 
sufficiently simple that an {\it analytic} solution for $p(r)$ can be
found, which corroborates the above remarks about not having a zero
at a finite $R$.
A suggestion for the student is to try a power-law Ansatz.

\subsection{The Fermi Gas EoS for Arbitrary Relativity}

In order to avoid the trap of the relativistic gas, one should find an
EoS for the non-interacting neutron Fermi gas which works 
for all values of the relativity parameter $x = k_F/m_n c$.
Taking a hint from the two polytropes, one can try to fit the
energy density as a function of pressure, each given as a transcendental
function of $k_F$, with the form
\begin{equation}
	\bar\epsilon(p) = A_{\rm NR}\bar{p}^{3/5} + A_{\rm R}\bar{p}
		\, .  \label{eq:NStarfitEoS}
\end{equation}
For low pressures the non-relativistic first term dominates over the second.
(The power in the relativistic term is changed from that in
Eq.\ (\ref{eq:WDfitEoS}).)
It is again useful to factor out an $\epsilon_0$ from both $\epsilon$ 
and $p$.
In this case, it is more natural to define it as
\begin{equation}
	\epsilon_0 = \frac{m_n^4 c^5}{(3 \pi^2 \hbar)^3} = 
		5.346 \times 10^{36} \ \frac{{\rm ergs}}{{\rm cm}^3} =
		0.003006 \ \frac{{\rm M}_\odot c^2}{{\rm km}^3}
		\, .  \label{eq:eps0fitEoS}
\end{equation}

Mathematica can easily create a table of exact $\bar\epsilon$ and $\bar p$ 
values as a function of $k_F$.
The dimensionless $A$-values can then be fit using its built-in fitting 
function.
From our efforts we found, to an accuracy of better than 1\% over most
of the range of $k_F$ \cite{lowkF},
\begin{equation}
	A_{\rm NR} = 2.4216 \ , \quad A_{\rm R} = 2.8663
		\, .  \label{eq:Avalues}
\end{equation}

We used the fitted functional form for $\bar\epsilon$ of 
Eq.\ (\ref{eq:NStarfitEoS}) in a Mathematica program similar to that 
for the neutron star based on the non-relativistic EoS.
With the $\epsilon_0$ of Eq.\ (\ref{eq:eps0fitEoS}) and choosing
$\alpha = R_0$ = 1.476 km, we obtain $\beta = 0.03778$.
Our results for a starting value of $\bar{p}(0) = 0.01$, clearly
in the relativistic regime, are
\begin{eqnarray}
	R &=& 15.0 \ , \quad M = 1.037 \, , \quad {\rm Newtonian\ equations} \\
	R &=& 13.4 \ , \quad M = 0.717 \, , \quad {\rm full\ TOV\ equation}
		\, .  \label{eq:RMresults0.01}
\end{eqnarray}
For this more massive star, the GR effects are significant (as should
be expected from the size of $G M / c^2 R$, about 10\% in this case).
Figure \ref{RMgraphs0.01} displays the differences.
\begin{figure}[tbp]
        \centering
	 \epsffile{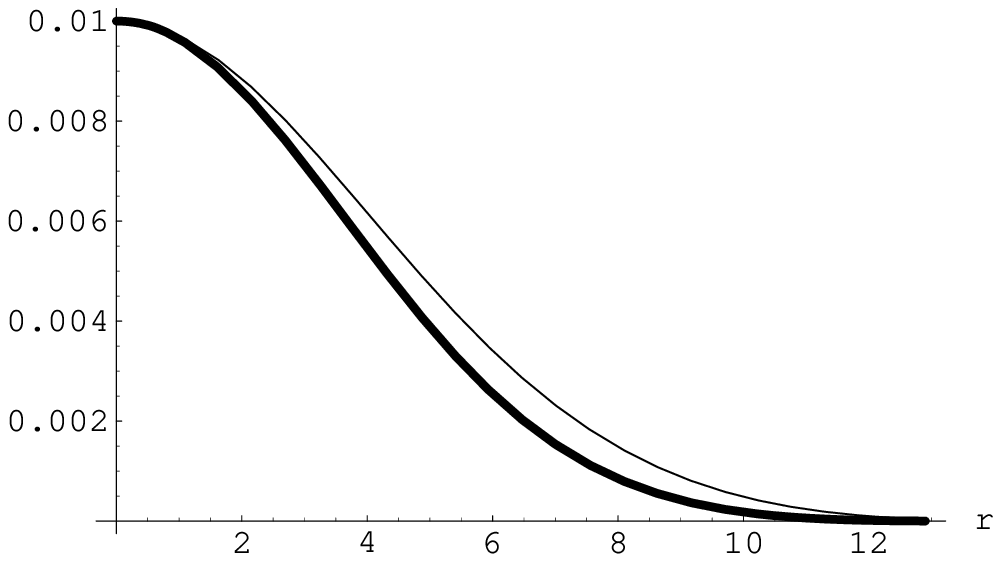}
        \epsffile{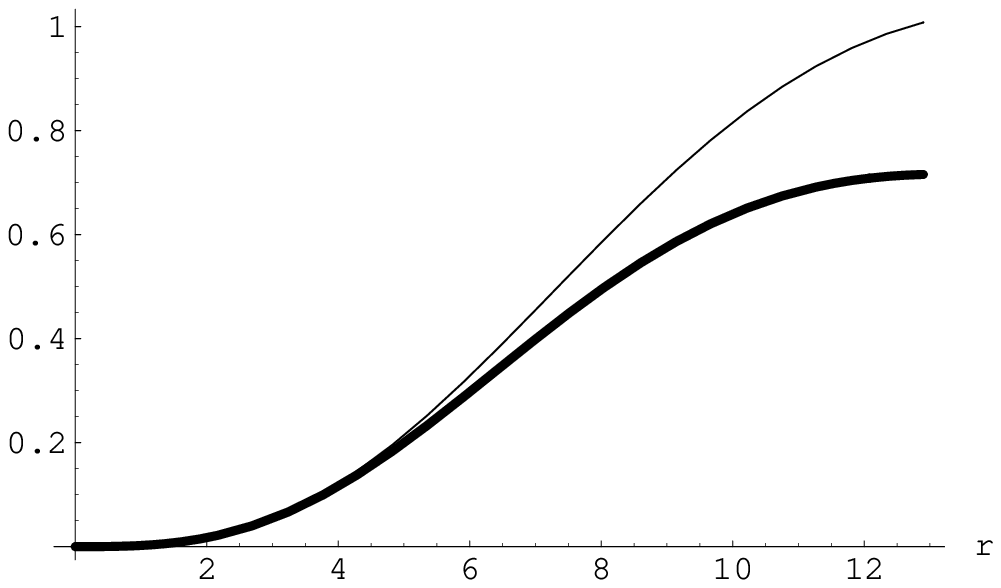}
        \caption{ $\bar{p}(r)$ and $\bar{\cal M}(r)$ ($r$ in km) 
        for a pure neutron
	star with central pressure $\bar{p}(0) = 0.01$, using a Fermi gas
	EoS fit valid for all values of $k_F$.  The thin curves are results 
	from the classical Newtonian structure equations, while the thick
	ones include general relativistic corrections.}
        \label{RMgraphs0.01}
\end{figure} 

It is now instructive to make a long run of calculations for a range of
$\bar{p}(0)$ values.
We display in Fig.\ \ref{RMplotFG} a (parametric) plot of $M$ and $R$ 
as they depend on the central pressure.
\begin{figure}[tbp]
        \centering
	$M(R)$\\
        \epsffile{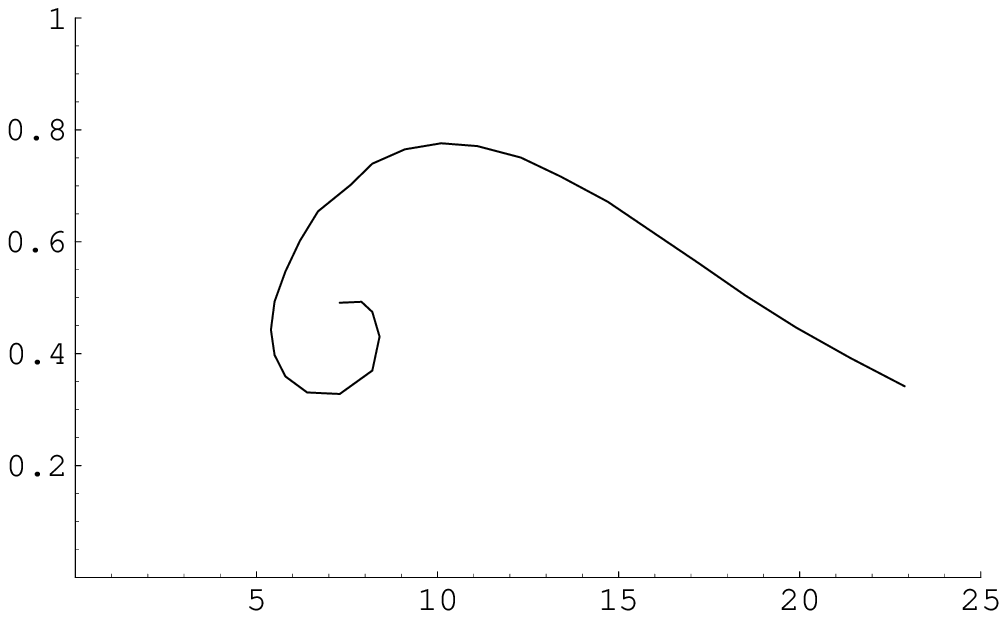}
        \caption{The mass $M$ (in M$_\odot$) and radius $R$ (in km) 
        for pure neutron
	stars, using a Fermi gas EoS.  The stars of low mass and large
	radius are solutions of the TOV equations for small values of
	central pressure $\bar{p}(0)$.  The stars to the right of the
	maximum at $R$ = 11 are stable, while those to the left will
	suffer gravitational collapse.}
        \label{RMplotFG}
\end{figure} 
The low-mass/large-radius stars are to the right in the graph and
correspond to small starting values of $\bar{p}(0)$.
As the central pressure increases, the total mass that the star can 
support increases.
And, the bigger the star mass, the bigger the gravitational attraction
which draws in the periphery of the star, making stars with smaller
radii.
That is, increasing $\bar{p}(0)$ corresponds to
``climbing the hill,'' moving upward and to the left in the diagram.

At about $\bar{p}(0)$ = 0.03 one gets to the top of the hill, 
achieving a maximum mass of about 0.8 ${\rm M}_\odot$ at a radius 
$R \approx 11$ km.
That maximum $M$ and its $R$ agree with Oppenheimer and Volkov's
seminal 1939 result for a Fermi gas EoS.

What about the solutions in Fig.\ \ref{RMplotFG} that are ``over the
hill,'' i.e., to the left of the maximum?
It turns out that these stars are unstable against gravitational collapse
into a black hole.
The question of stability, however, is a complicated issue \cite{WbrgStable},
perhaps too difficult for a student at this level.
The fact that things collapse to the left of the maximum, however, means
that one probably shouldn't worry too much about the peculiar curly-cue
tail to the $M$-$R$ curve in the figure.
It appears to be an artifact for {\it very} large values of $\bar{p}(0)$,
also seen in other calculations, even though it is especially prominent
for this Fermi gas EoS.

\subsection{Why Is There a Maximum Mass?}

On general grounds one can argue that cold compact objects such as white 
dwarfs and neutron stars must possess a limiting mass beyond which stable 
hydrostatic configurations are not possible. 
This limiting mass is often called the maximum mass of the object and was 
briefly mentioned in the discussion at the end of Sec.\ 2.2 and that 
relating to Fig. 6. 
In what follows, we outline the general argument.

The thermal component of the pressure in cold stars is by definition 
negligible. 
Thus, variations in both the energy density and pressure are only caused 
by changes in the density. 
Given this simple observation, let us examine why we expect a maximum 
mass in the Newtonian case. 

Here, an increase in the density results in a proportional increase in 
the energy density.  
This results in a corresponding increase in the gravitational 
attraction. 
To balance this, we require that the increment in pressure is large enough.
However, the rate of change of pressure with respect to  energy 
density is related to the speed of sound (see Sec.\ 6.3).
In a purely Newtonian world, this is in principle unbounded. 
However, the speed of all propagating signals cannot exceed the speed of 
light. 
This then puts a bound on the pressure increment associated with changes 
in density. 

Once we accept this bound, we can safely conclude that all cold compact 
objects will eventually run into the situation in which any increase in 
density will result in an additional gravitational attraction that cannot 
be compensated for by the corresponding increment in pressure. 
This leads naturally to the existence of a limiting mass for the star. 

When we include general relativistic corrections, as discussed in 
Sec.\ 2.2 earlier, they act to ``amplify'' gravity.
Thus we can expect the maximum mass to occur at a somewhat lower mass than in 
the Newtonian case.

\section{Neutron Stars with Protons and Electrons, Fermi Gas EoS}

As mentioned at the beginning of the last section, neutron stars are not
made only of neutrons.  
There must also be a small fraction of protons and electrons present.
The reason for this is that a free neutron will undergo a weak decay,
\begin{equation}
	n \rightarrow p + e^- + \bar\nu_e \ ,  \label{eq:ndecay}
\end{equation}
with a lifetime of about 15 minutes.
So, there must be something that prevents this decay in the case
of the star, and that is the presence of the protons and electrons.

The decay products have low energies ($m_n - m_p - m_e$ = 0.778 MeV),
with most of that energy being carried away by the light electron and 
(nearly massless) neutrino \cite{nucooling}.
If all the available low-energy levels for the decay proton are already
filled by the protons already present, then the Pauli exclusion principle
takes over and prevents the decay from taking place.

The same might be said about the presence of the electrons, but in any
case the electrons must be present within the star to cancel the positive 
charge of the protons.
A neutron star is electrically neutral.
We saw earlier that the number density of a particle species is fixed
in terms of that particle's Fermi momentum [see Eq.\ (\ref{eq:nofkF})].
Thus equal numbers of electrons and protons implies that
\begin{equation}
	k_{F,p} = k_{F,e} \, .  \label{eq:kFpeqkFe}
\end{equation}

In addition to charge neutrality, we also require weak interaction
equilibrium, i.e., as many neutron decays [Eq.\ (\ref{eq:ndecay})] taking
place as electron capture reactions, $p + e^- \rightarrow n + \nu_e$.
This equilibrium can be expressed in terms of the chemical potentials
for the three particle species,
\begin{equation}
	\mu_n = \mu_p + \mu_e \, .  \label{eq:wkeqbm}
\end{equation}
We already defined the chemical equilibrium for a particle in Sec.\ 3.2
after Eq.\ (\ref{eq:pressureDef}),
\begin{equation}
	\mu_i(k_{F,i}) = \frac{d \epsilon}{d n}  = 
		(k_{F,i}^2 + m_i^2)^{1/2} \, , \quad i = n, p, e
	\, .  \label{eq:chempot}
\end{equation}
where, for the time being, we have set $c = 1$ to simplify the 
equations somewhat.
(The student is urged to prove the right-hand equality.)
From Eqs.\ (\ref{eq:kFpeqkFe}), (\ref{eq:wkeqbm}), and (\ref{eq:chempot}) 
we can find a constraint determining $k_{F,p}$ for a given $k_{F,n}$,
\begin{equation}
	 (k_{F,n}^2 + m_n^2)^{1/2} - (k_{F,p}^2 + m_p^2)^{1/2} -
		(k_{F,p}^2 + m_e^2)^{1/2} = 0
		\, .  \label{eq:kFpconstraint}
\end{equation}
While an ambitious algebraist can probably solve this equation for
$k_{F,p}$ as a function of $k_{F,n}$, we were somewhat lazy and let 
Mathe\-matica do it, finding
\begin{eqnarray}
	k_{F,p}(k_{F,n}) &=& \frac{[(k_{F,n}^2 + m_n^2 - m_e^2)^2 
		- 2 m_p^2 (k_{F,n}^2 + m_n^2 + m_e^2) + m_p^4]^{1/2}}
		{2 (k_{F,n}^2 + m_n^2)^{1/2}} \label{eq:kFpfromkFn} \\
	&\approx& \frac{k_{F,n}^2 + m_n^2 - m_p^2}
		{2 (k_{F,n}^2 + m_n^2)^{1/2}} \ 
		{\rm\ \ for\ } \frac{m_e}{k_{F,n}} \rightarrow 0 \, .
\end{eqnarray}

The total energy density is the sum of the individual energy
densities,
\begin{equation}
	\epsilon_{\rm tot} = \sum_{i=n,p,e} \epsilon_i
		\, ,  \label{eq:epstot}
\end{equation}
where
\begin{equation}
	\epsilon_i(k_{F,i}) = \int_0^{k_{F,i}} (k^2 + m_i)^{1/2} k^2 dk
		= \epsilon_0 \, \bar\epsilon_i(x_i,y_i) \, ,  
		\label{eq:epssubi} 
\end{equation}
and, as before \cite{putbackcs},
\begin{eqnarray}
	\epsilon_0 &=& m_n^4 /3 \pi^2 \hbar^3 \, , \\
	\bar\epsilon_i(x_i,y_i) &=& 
		\int_0^{x_i} (u^2 + y_i^2)^{1/2} u^2 du \, , \\
	x_i &=& k_{F,i}/m_i \, , \  y_i = m_i/m_n \, .
		\label{eq:eps0again} 
\end{eqnarray}
The corresponding total pressure is
\begin{eqnarray}
	p_{\rm tot} &=& \sum_{i=n,p,e} p_i \, ,  \label{eq:prestot} \\
	p_i(k_{F,i}) &=& \int_0^{k_{F,i}} (k^2 + m_i)^{-1/2} k^4 dk
		= \epsilon_0 \, \bar{p}_i(x_i,y_i) \, , \\
	\bar{p}_i(x_i,y_i) &=& 
		\int_0^{x_i} (u^2 + y_i^2)^{-1/2} u^4 du \, .
\end{eqnarray}

Using Mathematica the (dimensionless) integrals can be expressed in
terms of log and $\sinh^{-1}$ functions of $x_i$ and $y_i$.
One can then generate a table of $\bar\epsilon_{\rm tot}$ versus 
$\bar{p}_{\rm tot}$ values for an appropriate range of $k_{F,n}$'s.
This, in turn, can be fitted to the same form of two terms
as before in Eq.\ (\ref{eq:NStarfitEoS}).
We found, this time, the coefficients to be
\begin{equation}
	A_{\rm NR} = 2.572 \ , \quad A_{\rm R} = 2.891
		\, .  \label{eq:npeAvalues}
\end{equation}
These coefficients are not much changed from those in Eq.\ (\ref{eq:Avalues})
for the pure neutron star.
Therefore, we expect that the $M$ versus $R$ diagram for this more realistic
Fermi gas model would not be much different from that in Fig. \ref{RMplotFG}.

\section{Introducing Nuclear Interactions}

Nucleon-nucleon interactions can be included in the EoS
(they are important) by constructing a simple model for the nuclear potential 
that reproduces the general features of (normal) nuclear matter.
In doing so we were much guided by the lectures of Prakash \cite{Prakash}.

We will use MeV and fm ($10^{-13}$ cm) as our energy and distance units
for much of this section, converting back to ${\rm M}_\odot$ 
and km later.  
We will also continue setting $c=1$ for now.
In this regard, the important number to remember for 
making conversions is $\hbar c = 197.3$ MeV-fm.
We will also neglect the mass difference between protons and neutrons,
labeling their masses as $m_N$.

The von Weiz\"acker mass formula \cite{massformula} for nuclides 
with $Z$ protons and $N$ neutrons gives, for
normal symmetric nuclear matter ($A = N+Z$ with $N=Z$), an equilibrium number 
density $n_0$ of 0.16 nucleons/fm$^3$.
For this value of $n_0$ the Fermi momentum is $k_F^0 = 263$ MeV/$c$
[see Eq.\ (\ref{eq:nofkF})].
This momentum is small enough compared with $m_N = 939$ MeV/c$^2$ to
allow a non-relativistic treatment of normal nuclear matter.
At this density, the average binding energy per nucleon, $BE = E/A - m_N$, 
is $-16$ MeV.
These are two physical quantities we definitely want our nuclear potential 
to respect, but there are two more that we'll need to fix the parameters 
of the model.

We chose one of these as the {\it nuclear compressibility}, $K_0$, to be
defined below.
This is a quantity which is not all that well established but is in the range
of 200 to 400 MeV.
The other is the so-called {\it symmetry energy} term, which, when $Z=0$,
contributes about 30 MeV of energy above the symmetric matter minimum at $n_0$.
(This quantity might really be better described as an asymmetry 
parameter, since it accounts for the energy that comes in when $N \neq Z$.)

\subsection{Symmetric Nuclear Matter}

We defer the case when $N \neq Z$, which is our main interest in this
paper, to the next sub-section.  
Here we concentrate on getting a good (enough) EoS for nuclear
matter when $N = Z$, or, equivalently, when the proton and neutron
number densities are equal, $n_n = n_p$.
The total nucleon density $n = n_n + n_p$.

We need to relate the first three nuclear quantities, $n_0$, $BE$, and 
$K_0$ to the energy density for symmetric nuclear matter, $\epsilon(n)$.
Here $n = n(k_F)$ is the nuclear density (at and away from $n_0$).
We will not worry in this section about the electrons that are present,
since, as was seen in the last section, its
contribution is small.
The energy density now will include the nuclear potential, $V(n)$,
which we will model below in terms of two simple functions with three 
parameters that are fitted to reproduce the above three nuclear quantities.
[The fourth quantity, the symmetry energy, will be used in the next
sub-section to fix a term in the potential which is proportional to 
$(N-Z)/A$.]

First, the average energy per nucleon, $E/A$, for symmetric 
nuclear matter is related to $\epsilon$ by
\begin{equation}
	E(n)/A = \epsilon(n)/n \, ,
\end{equation}
which includes the rest mass energy, $m_N$ and has dimensions of MeV.
As a function of $n$, $E(n)/A - m_N$ has a minimum at $n = n_0$ with a depth
$BE = -16$ MeV.  This minimum occurs when
\begin{equation}
	\frac{d}{d n}\left(\frac{E(n)}{A}\right) = 
		\frac{d}{d n}\left(\frac{\epsilon(n)}{n}\right)
		= 0  {\ \rm\ at\ } n = n_0 \, . \label{eq:n0constraint}
\end{equation}
This is one constraint of the parameters of $V(n)$.  
Another, of course, is the binding energy,
\begin{equation}
	\frac{\epsilon(n)}{n} - m_N = BE  {\ \rm\ at\ } n = n_0 \, . 
		\label{eq:BEconstraint}
\end{equation}
The positive curvature at the bottom of this valley is related
to the nuclear (in)com\-press\-ibility by \cite{compressibility}
\begin{equation}
	K(n) = 9 \frac{d p(n)}{d n} = 
		9 \left[ n^2 \frac{d^2}{d n^2} 
				\left( \frac{\epsilon}{n} \right)
			+ 2n \frac{d}{d n} 
				\left( \frac{\epsilon}{n} \right) \right]
	  	 \, , \label{eq:Kconstraint}
\end{equation}
using Eq.\ (\ref{eq:pressureDef}), which defines the pressure in terms of 
the energy density.
At $n = n_0$ this quantity equals $K_0$.
(The factor of 9 is a historical artifact from the convention originally
defining $K_0$.)

(Question: why does one {\it not} have to calculate the pressure at 
$n = n_0$?)

The $N=Z$ potential in $\epsilon(n)$ we will model as \cite{Prakash}
\begin{equation}
	\frac{\epsilon(n)}{n} = m_N + 
		\frac{3}{5} \frac{\hbar^2 k_F^2}{2 m_N} + 
		\frac{A}{2} u + \frac{B}{\sigma + 1} u^\sigma
	        \, , \label{eq:epsbyn}
\end{equation}
where $u = n/n_0$ and $\sigma$ are dimensionless and $A$ and $B$ have
dimensions of MeV.  
The first term represents the rest mass energy and the second the
average kinetic energy per nucleon.
[These two terms are leading in the non-relativistic limit of the
nucleonic version of Eq.\ (\ref{eq:electroneps}).]
For $k_F(n_0) = k_F^0$ we will abbreviate the kinetic energy term as 
$\left<E_F^0\right>$, which evaluates to 22.1 MeV.
The kinetic energy term in Eq.\ (\ref{eq:epsbyn}) can be better
written as $\left<E_F^0\right> u^{2/3}$.

From the above three constraints, 
Eqs.\ (\ref{eq:n0constraint})-(\ref{eq:Kconstraint}), and noting
that $u = 1$ at $n = n_0$,
we get three equations for the parameters $A$, $B$, and $\sigma$:
\begin{eqnarray}
	\left<E_F^0\right> + \frac{A}{2} + \frac{B}{\sigma + 1} &=& BE \, , 
		\label{eq:sigparamEqn}\\
	\frac{2}{3}\left<E_F^0\right>+\frac{A}{2} + 
			\frac{B\sigma}{\sigma+1} &=& 0\, , 
		\label{eq:BparamEqn}\\
	\frac{10}{9}\left<E_F^0\right> + A + B\sigma &=& \frac{K_0}{9} \, .
		\label{eq:AparamEqn}
\end{eqnarray}
Solving these equations (which we found easier to do by hand than
with Mathematica), one finds
\begin{eqnarray}
	\sigma &=& \frac{K_0 + 2\left<E_F^0\right>}
			{3\left<E_F^0\right> - 9 BE} \, , \\
	B &=& \frac{\sigma+1}{\sigma-1} 
		\left[ \frac{1}{3}\left<E_F^0\right> - BE \right] \, ,\\
	A&=& BE - \frac{5}{3}\left<E_F^0\right> - B \, .
\end{eqnarray}
Numerically, for $K_0 = 400$ MeV (which is perhaps a high value),
\begin{equation}
	A = -122.2 {\ \rm MeV}, \quad B = 65.39 {\ \rm MeV}, \quad  
	\sigma = 2.112 \, . \label{eq:params400}
\end{equation}
Note that $\sigma > 1$, a point we will come back to below, since it
violates a basic principle of physics called ``causality.''

The student can try other values of $K_0$ to see how the parameters
$A$, $B$, and $\sigma$ change.
More interesting is to see how the interplay between the $A$- and
$B$-terms gives the valley at $n = n_0$.
Figure \ref{EbyAofn} shows $E/A - m_N$ as a function of $n$ using the
parameters of Eq.\ (\ref{eq:params400}).
\begin{figure}[tbp]
        \centering
	$E/A - m_N$\\
        \epsffile{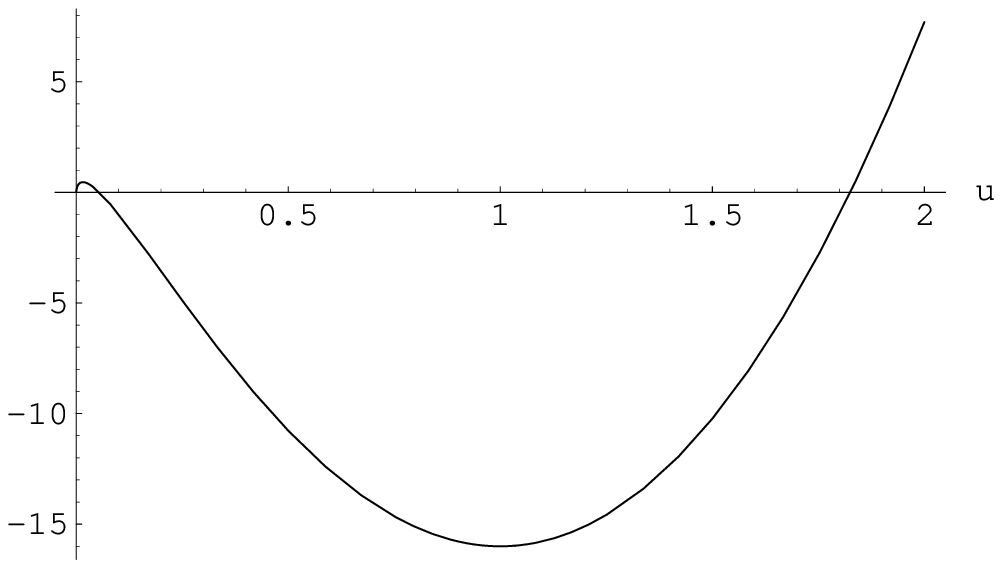}
        \caption{The average energy per nucleon, less its rest mass,
	as a function of $u = n/n_0$ (in MeV).  The position of the minimum is
	at $n = n_0 = 0.16$ fm$^{-3}$, its depth there is $BE = -16$ MeV,
	and its curvature (second derivative) there corresponds to
	nuclear compressibility $K_0 = 400$ MeV.}
	\label{EbyAofn}
\end{figure}
We would hope the student notices the funny little positive bump
in this plot near $n=0$ and sorts out the reason for its occurrence.

Given $\epsilon(n)$ from Eq.\ (\ref{eq:epsbyn}) one can find the
pressure using Eq.\ (\ref{eq:pressureDef}), 
\begin{equation}
	p(n) = n^2 \frac{d}{d n} \left( \frac{\epsilon}{n} \right) =
		n_0 \left[ \frac{2}{3}\left<E_F^0\right> u^{5/3} + 
			 \frac{A}{2} u^2 + 
			 \frac{B\sigma}{\sigma+1} u^{\sigma+1} \right]
		 \, . \label{eq:pressofnNeqZ}
\end{equation}
For the parameters of Eq.\ (\ref{eq:params400}) its dependence on
$n$ is shown in Fig.\ \ref{presofn}.
\begin{figure}[tbp]
        \centering
	$p(u)$\\
        \epsffile{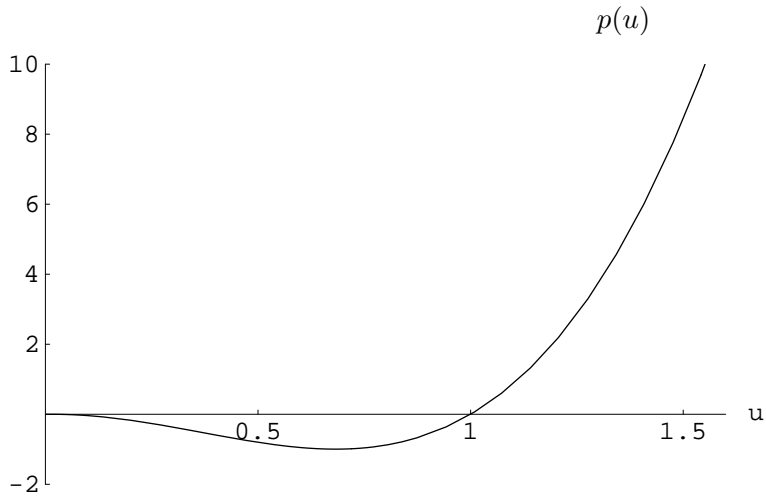}
        \caption{The pressure for symmetric nuclear matter
	as a function of $u = n/n_0$.  The student should ask what
	it means when the pressure is negative and why it is 0 at $u=1$.}
	\label{presofn}
\end{figure}
On first seeing this graph, the student should wonder why 
$p(u=1) = p(n_0) = 0$.  
And, what is the meaning of the negative values for pressure below $u=1$?
(Hint: what is ``cavitation''?)

So, if this $N=Z$ case were all we had for the nuclear EoS, a plot of
$\epsilon(n)$ versus $p(n)$ would only make sense for $n \geq n_0$.
Such a plot looks much like a parabola opening to the right for the 
range $0 < u < 3$.
At very large values of $u$, however, $\epsilon \approx p/3$, as it
should for a relativistic nucleon gas (see Section 4.2).
We don't pursue this symmetric nuclear matter EoS further since our
interest is in the case when $N \gg Z$ \cite{RHIC}.

\subsection{Non-Symmetric Nuclear Matter}

We continue following Prakash's notes \cite{Prakash} closely.
Let us  represent the neutron and proton densities in 
terms of a parameter $\alpha$ as
\begin{equation}
	n_n = \frac{1 + \alpha}{2}\,n \, , \quad
	n_p = \frac{1 - \alpha}{2}\,n 
		 \, . \label{eq:newalphadef}
\end{equation}
This $\alpha$ is not to be confused with the constant defined in 
Eq.\ (\ref{eq:alphadef}).
For pure neutron matter $\alpha = 1$.
Note that
\begin{equation}
	\alpha = \frac{n_n - n_p}{n} = \frac{N-Z}{A}
		 \, , \label{eq:alphaIsNminusZ}
\end{equation}
so we can expect that the isospin-symmetry-breaking interaction
is proportional to $\alpha$ (or some power of it).
An alternative notation is in terms of the fraction of protons in
the star,
\begin{equation}
	x = \frac{n_p}{n} = \frac{1 - \alpha}{2}
		 \, . \label{eq:pfracdef}
\end{equation}
We now consider how the energy density changes from the symmetric case 
discussed above, where $\alpha = 0$ (or $x = 1/2$).

First, there are contributions to the kinetic energy part of $\epsilon$
from both neutrons and protons,
\begin{eqnarray}
	\epsilon_{KE}(n,\alpha) &=& \frac{3}{5} 
		\frac{k_{F,n}^2}{2 m_N}\, n_n + 
		\frac{3}{5} \frac{k_{F,p}^2}{2 m_N}\, n_p  
		\nonumber \\
	&=& n \left<E_F\right> \frac{1}{2} \left[ 
			\left(1+\alpha\right)^{5/3} + 
			\left(1-\alpha\right)^{5/3} \right] \,  ,
\end{eqnarray}
where
\begin{equation}
	\left<E_F\right> = \frac{3}{5} \frac{\hbar^2}{2 m_N}
			\left( \frac{3 \pi^2 n}{2} \right)^{2/3}
		 \,  \label{eq:EFdef}
\end{equation}
is the mean kinetic energy of symmetric nuclear matter at density $n$.
For $n=n_0$ we note that $\left<E_F\right> = 3 \left<E_F^0\right>/5$
[see Eq.\ (\ref{eq:epsbyn})].
For non-symmetric matter, $\alpha \neq 0$, the excess kinetic energy is
\begin{eqnarray}
	\Delta\epsilon_{KE}(n,\alpha) &=& \epsilon_{KE}(n,\alpha) - 
			\epsilon_{KE}(n,0) \nonumber \\
		&=& n \left<E_F\right> \left\{ \frac{1}{2} \left[ 
			\left(1+\alpha\right)^{5/3} + 
			\left(1-\alpha\right)^{5/3} \right]
				- 1 \right\} \nonumber \\
		&=& n \left<E_F\right> \left\{ 2^{2/3} \left[ 
			\left(1-x\right)^{5/3} + x^{5/3} \right]
				- 1 \right\}
		 \, . \label{eq:delepsKE}
\end{eqnarray}
For pure neutron matter, $\alpha = 1$,
\begin{equation}
	\Delta\epsilon_{KE}(n,\alpha) = n \left<E_F\right>
		\left( 2^{2/3} - 1 \right) \, .
\end{equation}
It is also useful to expand to leading order in $\alpha$,
\begin{eqnarray}
	\Delta\epsilon_{KE}(n,\alpha) &=& n \left<E_F\right> \frac{5}{9} 
		\alpha^2 \left( 1 + \frac{\alpha^2}{27} + \cdots \right) \\
		&=&  n \, E_F \, \frac{\alpha^2}{3}\,
		 \left( 1 + \frac{\alpha^2}{27} + \cdots \right)
		\, .
\end{eqnarray}
Keeping terms to order $\alpha^2$ is evidently good enough for most
purposes.
For pure neutron matter, the energy per particle (which, recall, is 
$\epsilon/n$) at normal density is 
$\Delta\epsilon_{KE}(n_0,1)/n_0 \approx 13$ MeV,
more than a third of the total bulk symmetry energy of 30 MeV, our fourth
nuclear parameter.

Thus the potential energy contribution to the bulk symmetry energy must
be 20 MeV or so.
Let us assume the quadratic approximation in $\alpha$ also works well enough 
for this potential contribution and write the total energy per particle as
\begin{equation}
	E(n,\alpha) = E(n,0) + \alpha^2 S(n) \, , \label{eq:Enalphadef}
\end{equation}
The isospin-symmetry breaking is proportional to $\alpha^2$, which
reflects (roughly) the pair-wise nature of the nuclear interactions.

We will assume $S(u)$, $u = n/n_0$, has the form
\begin{equation}
	S(u) = (2^{2/3} - 1) \frac{3}{5} \left<E_F^0\right> 
		\left( u^{2/3} - F(u) \right) + S_0 F(u)\, .
		\label{SofuDef}
\end{equation}
Here $S_0 = 30$ MeV is the bulk symmetry energy parameter. 
The function $F(u)$ must satisfy $F(1) = 1$ [so that $S(u=1) = S_0$]
and $F(0) = 0$ [so that $S(u=0) = 0$; no matter means no energy].
Besides these two constraints there is, from what we presently know, a lot 
of freedom in what one chooses for $F(u)$.  
We will make the simplest possible choice here, namely,
\begin{equation}
	F(u) = u \, , \label{FofuLinear}
\end{equation}
but we encourage the student to try other forms satisfying the conditions
on $F(u)$, such as $\sqrt u$, to see what difference it makes.

Figure \ref{EperAneut} shows the energy per particle for pure neutron 
matter, $E(n,1) - m_N$, as a function of $u$ for the parameters of 
Eq.\ (\ref{eq:params400}) and $S_0 = 30$ MeV.
\begin{figure}[tbp]
        \centering
	$E(n,\alpha=1) - m_N$\\
        \epsffile{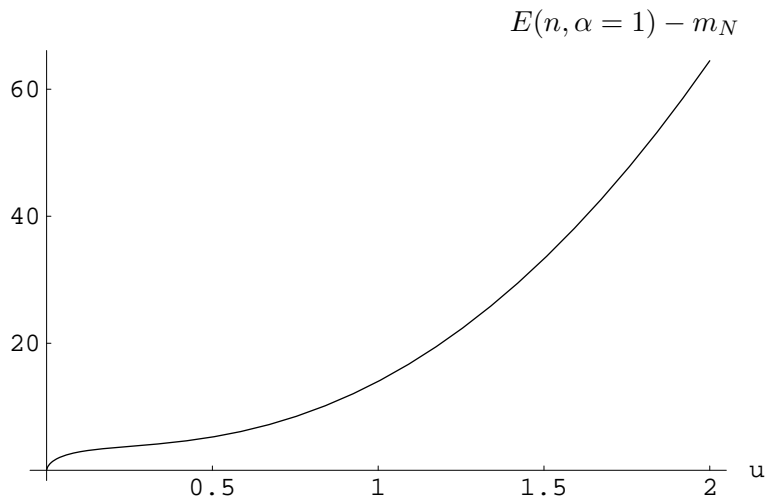}
        \caption{The average energy per neutron (less its rest mass), 
	in MeV, for pure neutron matter, as a function of $u = n/n_0$.  
	The parameters for this curve are for a nuclear compressibility 
	$K_0$ of 400 MeV.}
	\label{EperAneut}
\end{figure}
In contrast with the $\alpha = 0$ plot in Fig.\ \ref{EbyAofn},
$E(n,1) \geq 0$ and is monotonically increasing.
The plot looks almost quadratic as a function of $u$.
The dominant term at large $u$ goes like 
$u^\sigma$, and $\sigma = 2.112$ (for this case).
However, one might have expected a {\it linear} increase instead.
We will return to this point in Sec.\ 6.3.

Given the energy density, $\epsilon(n,\alpha) = n_0 u E(n,\alpha)$,
the corresponding pressure is, from Eq.\ (\ref{eq:pressureDef}),
\begin{eqnarray}
	p(n,x) &=& u \frac{d}{du}\epsilon(n,\alpha) - 
			\epsilon(n,\alpha) \nonumber \\
		&=& p(n,0) + n_0 \alpha^2 \left[
			\frac{2^{2/3}-1}{5}\left<E_F^0\right>
				\left(2 u^{5/3} - 3 u^2\right) +
			S_0 u^2 \right]  \, , \label{eq:presnpdef}
\end{eqnarray}
where $p(n,0)$ is defined by Eq.\ (\ref{eq:pressofnNeqZ}).
Figure \ref{pepsnpPlot} shows the dependence of the pure neutron 
$p(n,1)$ and $\epsilon(n,1)$
on $u = n/n_0$, ranging from 0 to 10 times normal nuclear density.
\begin{figure}[tbp]
        \centering
	$p$ and $\epsilon$ versus $u$\\
        \epsffile{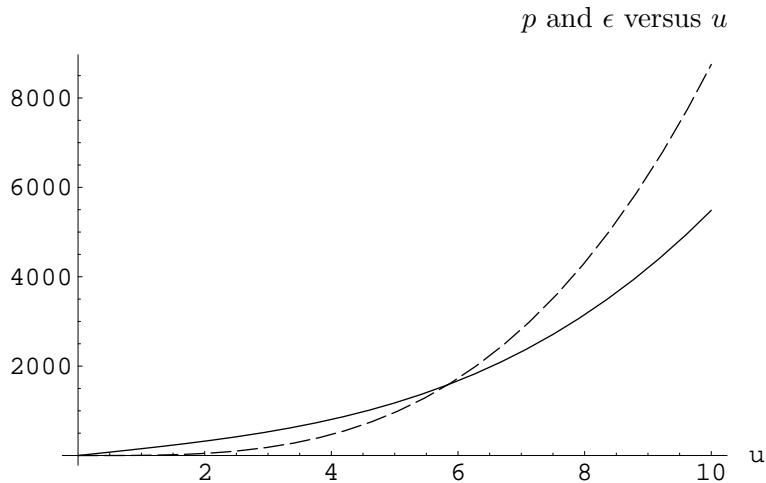}
        \caption{The pressure (dashed curve) and energy density (solid)
	for pure neutron matter, as a function of $u = n/n_0$.
	Units for the $y$-axis are MeV/fm$^3$.  
	This curve uses parameters based on a nuclear compressibility 
	$K_0 = 400$ MeV.}
	\label{pepsnpPlot}
\end{figure}
Both functions increase smoothly and monotonically from $u = 0$.
We hope the student would wonder why the pressure becomes greater
than the energy density around $u = 6$.
Why doesn't it go like a relativistic nucleon gas, $p = \epsilon/3$?
(Hint: check the assumptions.)

One can now look at the EoS, i.e., the dependance of $p$ on $\epsilon$ 
(the points in Fig.\ \ref{eosfit}).
\begin{figure}[tbp]
        \centering
	$p$ versus $\epsilon$\\
        \epsffile{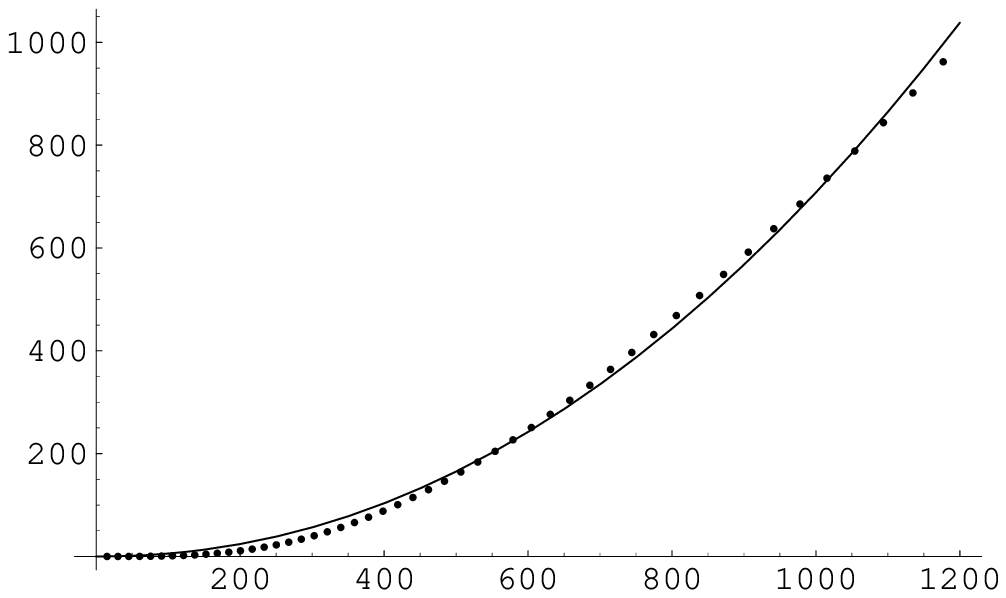}
        \caption{The equation of state for pure neutron matter ($\alpha=1$), 
	i.e., the dependence of pressure ($y$-axis) versus energy density 
	($x$-axis).
	Units for both axes are MeV/fm$^3$, and the nuclear compressibility 
	in this case is $K_0 = 400$ MeV.
	The points are values calculated directly from 
	Eq.\ (\ref{eq:Enalphadef}), multiplied by $n$, and 
	Eq.\ (\ref{eq:presnpdef}), while the solid curve is a fit to 
	these points given in Eqs.\ (\ref{eq:eosfitform}) and
	(\ref{eq:eosfitparams}).}
	\label{eosfit}
\end{figure}
The pressure is smooth, non-negative, and monotonically increasing as a
function of $\epsilon$.
In fact it looks almost quadratic over this energy range ($0\leq u\leq 5$).
This suggests that it might be reasonable to see if one can make a simple, 
polytropic fit.
If we try that using a form
\begin{equation}
	p(\epsilon) = \kappa_0 \, \epsilon^\gamma \, , \label{eq:eosfitform}
\end{equation}
we find the fit shown in Fig.\ \ref{eosfit} as the solid curve with
\begin{equation}
	\kappa_0 = 3.548 \times 10^{-4} \, , \quad \gamma = 2.1
	\, , \label{eq:eosfitparams}
\end{equation}
where $\kappa_0$ has appropriate units so that $p$ and $\epsilon$ are
in MeV/fm$^3$.  
(We simply guessed and set $\gamma$ to that value.)

This polytrope can now be used in solving the TOV equation for a pure
neutron star with nuclear interactions.
Alternatively, one might solve for the structure by using the functional
forms from Eq.\ (\ref{eq:Enalphadef}), multiplied by $n$, and 
Eq.\ (\ref{eq:presnpdef}) directly. 
We defer that for a bit, since it would be a good idea to first find
an EoS which doesn't violate causality, a basic tenet of special relativity.

\subsection{Does the Speed of Sound Exceed That of Light?}

What is the speed of sound in nuclear matter?
Starting from the elementary formula for the square of the speed of sound 
in terms of the bulk modulus \cite{young},
one can show that
\begin{equation}
	\left( \frac{c_s}{c} \right)^2 = \frac{B}{\rho c^2} =
		\frac{d p}{d \epsilon} =
		\frac{dp/dn}{d\epsilon/dn} \, . \label{eq:soundspeed}
\end{equation}
To satisfy relativistic causality we must
require that the sound speed does not exceed that of light.
This can happen when the density becomes very
large, i.e., when $u\rightarrow\infty$.
For the simple model of nuclear interactions presented in the last section,
the dominant terms at large $u$ in $p$ and $\epsilon$ are those going like 
$u^{\sigma +1}$. 
Thus, from Eq.\ (\ref{eq:Enalphadef}), multiplied by $n$, and 
Eq.\ (\ref{eq:presnpdef}), we see that 
\begin{equation}
	\left( \frac{c_s}{c} \right)^2 =
		\frac{dp/dn}{d\epsilon/dn} \rightarrow \sigma = 2.11
		\,  \label{eq:soundgt1}
\end{equation}
for the parameters of Eq.\ (\ref{eq:params400}), and indeed for any set
of parameters with $K_0$ greater than about 180 MeV.

One can recover causality (i.e., speeds of sound less than light) by assuring 
that both $\epsilon(u)$ and $p(u)$ grow no faster than $u^2$.
There must still be an interplay between the $A$- and $B$-terms in
the nuclear potential, but one simple way of doing this is to modify
the $B$-term by introducing a fourth parameter $C$ so that, for symmetric
nuclear matter ($\alpha = 0$),
\begin{equation}
	V_{\rm Nuc}(u,0) = \frac{A}{2} u + \frac{B}{\sigma+1} \
		\frac{u^\sigma}{1 + C u^{\sigma-1}}
		\, . \label{eq:modVnuc}
\end{equation}
One can choose $C$ small enough so that the effect of the denominator
only becomes appreciable for very large $u$.
The presence of the denominator would modify and complicate the constraint
equations for $A$, $B$, and $\sigma$ from those given in 
Eqs.\ (\ref{eq:sigparamEqn})--(\ref{eq:AparamEqn}).
However, for small $C$, which can be chosen as one wishes, the values
for the other parameters should not be much changed from those, say,
in Eq.\ (\ref{eq:params400}).
Thus, with a little bit of trial and error, one can simply readjust
the $A$, $B$, and $\sigma$ values to put the minimum of $E/A - m_N$ at 
the right position ($n_0$) and depth ($BE$), hoping that the resulting
value of the (poorly known) compressibility $K_0$ remains sensible.

In our calculations we chose $C = 0.2$ and started the hand search with
the $K_0 = 400$ MeV parameters in Eq.\ (\ref{eq:params400}).
We found that, by fiddling only with $B$ and $\sigma$, we could
re-fit $n_0$ and $B$ with only small changes,
\begin{equation}
	B = 65.39 \rightarrow 83.8 {\rm\ MeV} \, , \quad 
	\sigma = 2.11 \rightarrow 2.37 \, , \label{eq:refitparams}
\end{equation}
somewhat larger than before.
For these new values of $B$ and $\sigma$, $A$ changes from -122.2 MeV to 
-136.7 MeV, and $K_0$ from 400 to 363.2 MeV.
That is, it remains a reasonable nuclear model.

One can now proceed as in the last section to get $\epsilon(n,\alpha)$,
$p(n,\alpha)$, and the EoS, $p(\epsilon,\alpha)$.
The results are not much different from those shown in the figures of
the previous sub-section.
This time we decided to live with a quadratic fit for the EoS for pure
neutron matter, finding
\begin{equation}
	p(\epsilon,1) = \kappa_0 \epsilon^2 \, , \quad
	\kappa_0 = 4.012 \times 10^{-4} \, . \label{refitEoS}
\end{equation}
This is not much different from before, Eq.\ (\ref{eq:eosfitparams}).
Somewhat more useful for solving the TOV equation is to express 
$\epsilon$ in terms of $p$,
\begin{equation}
	\epsilon(p) = \left(p/\kappa_0)\right)^{1/2} \, .
	\label{eq:causalepsofp}
\end{equation}

\subsection{Pure Neutron Star with Nuclear Interactions}

Having laid all this groundwork, the student can now proceed to solve
the TOV equations as before for a pure neutron star, using the
fit for $\epsilon(p)$ found in the previous sub-section.
It is, once again, useful to convert from the units of MeV/fm$^3$ to
ergs/cm$^3$ to
${\rm M}_\odot$/km$^3$ and dimensionless $\bar{p}$ and $\bar\epsilon$.
By now the student has undoubtedly grown quite accustomed to that procedure.
\begin{equation}
	\bar\epsilon(\bar{p}) = (\kappa_0 \epsilon_0)^{-1/2} \bar{p}^{1/2}
		= A_0 \bar{p}^{1/2} \, , \quad 
		A_0 = 0.8642 \, , \label{eq:nuclepsbar}
\end{equation} 
where this time we defined 
\begin{equation}
	\epsilon_0 = \frac{m_n^4 c^5}{3 \pi^2 \hbar^3} \, . 
\end{equation}
With this, the constant $\alpha$ that occurs on the right-hand side
of the TOV equation, Eq.\ (\ref{eq:DEpresDimless}), is 
$\alpha = A_0 R_0 = 1.276$ km.
The constant for the mass equation, Eq.\ (\ref{eq:DEcurlyMDimless}), 
is $\beta = 0.03265$, again in units of 1/km$^3$.

Now proceeding as before, one can solve the coupled TOV equations for
$\bar{p}(r)$ and $\bar{\cal M}(r)$ for various initial central pressures,
$\bar{p}(0)$.
We don't exhibit here plots of the solutions, as they look very similar
to those for the Fermi gas EoS, Fig.\ \ref{RMgraphs0.01}.

More interesting is to solve for a range of initial $\bar{p}(0)$'s,
generating, as before, a mass $M$ versus radius $R$ plot which now includes  
nucleon-nucleon interactions (Fig.\ \ref{MRplotVN}).
\begin{figure}[tbp]
        \centering
	$M(R)$\\
        \epsffile{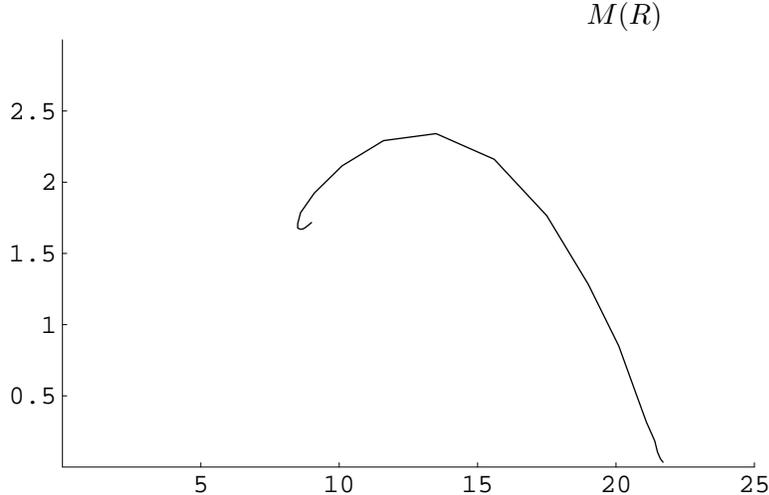}
        \caption{The mass $M$ and radius $R$ for pure neutron stars
	using an EoS which contains nucleon-nucleon interactions.
	Only those stars to the right of the maximum are stable
	against gravitational collapse.
	Compare this graph with that in Fig.\ \ref{RMplotFG} which is based
	on a non-interacting Fermi gas model for the EoS.}
        \label{MRplotVN}
\end{figure} 
The effect of the nuclear potential is enormous, on comparing with the 
Fermi gas model predictions for $M$ vs. $R$ shown in Fig.\ \ref{RMplotFG}.
The maximum star mass this time is about 2.3 ${\rm M}_\odot$, rather 
than 0.8 ${\rm M}_\odot$.
The radius for this maximum mass star is about 13.5 km, somewhat larger
than the Fermi gas model radius of 11 km.
The large value of maximum $M$ is a reflection of the large value of
nuclear (in)compressibility $K_0 = 363$ MeV.
The more incompressible something is, the more mass it can support.
Had we fit to a smaller
value of $K_0$ we would have gotten a smaller maximum mass.

\subsection{What About a Cosmological Constant?}

We do not know (either) if there is one, but there are definite indications
that a great part of the make-up of our universe is something called
``Dark Energy'' \cite{darkenergy}.
This conclusion comes about because
we have recently learned that something, at the present time, is causing
the universe to be accelerating, instead of slowing down (as would be 
expected after the Big Bang).

One way (of several) to interpret this dark energy is as Einstein's
cosmological constant, which contributes a term $\Lambda g_{\mu \nu}$
to the right-hand side of Einstein's field equation,
the basic equation of general relativity.
The most natural value for $\Lambda$ would be zero, but that may not
be the way the world is.
If $\Lambda$ {\it is} non-zero, it is nonetheless surprisingly small.

What would the effect of a non-zero cosmological constant be for the 
structure of a neutron star?
It turns out that the only modification to the TOV equation \cite{Pauchy} 
is in the correction factor
\begin{equation}
	\left[1 + \frac{4 \pi r^3 p(r)} {{\cal M}(r) c^2} \right] \rightarrow
	\left[1 + \frac{4 \pi r^3 p(r)} {{\cal M}(r) c^2} -
		\frac{\Lambda r^3}{2 G {\cal M}(r)}     \right]
		\, . \label{TOVwLambda}
\end{equation}
So, we encourage the student to, first, understand the units of $\Lambda$
and then to see what values for it might affect the structure of a
typical neutron star.

\section{Conclusions}

The materials we have described in this paper would be quite suitable as
an undergraduate thesis or special topics course accessible to a junior 
or senior physics major.
It is a topic rich in the subjects the student will have covered in
his or her courses, ranging from thermodynamics to quantum statistics
to nuclear physics.

The major emphasis in such a project is on constructing a (simple) 
equation of state.
This is needed to be able to solve the non-linear structure equations.
Solving those equations numerically, of course, develops the student's
computational skills.
Along the way, however, he or she will also learn some of the lore
regarding degenerate stars, e.g., white dwarfs and neutron stars.
And, in the latter case, the student will also come to appreciate the
relative importance of special and general relativity.

\section{Acknowledgments}  

We thank M.\ Prakash, T.\ Buervenich, and Y.-Z.\ Qian for their
helpful comments and for reading drafts of this paper. We would also
like to acknowledge comments and suggestions made by an anonymous
referee.  This research is supported in part by the Department of Energy
under contract W-7405-ENG-36.



\begin{thebibliography}{9}
\bibitem{CGS} 
We apologize to readers who are enthusiasts of SI units, but the first author
was raised on CGS units.  Actually, we strongly feel that by the time a
physics student is at this level, he or she ought to be comfortable
in switching from one system of units to another.
\bibitem{notCoul}
Despite the appearance of the $4\pi\epsilon_0$, the astute student will 
not be lulled into thinking that this factor has anything to do with a 
Coulomb potential or the dielectric constant of the vacuum.
\bibitem{monofall}
Note that the right-hand side of Eq.\ (\ref{eq:DEpresDimless}) is negative
(for positive $\bar p$), so $\bar p(r)$ must fall monotonically from
$\bar p(0)$.
\bibitem{hint}
We leave this for the student to figure out, except for the following
hint: Use an if statement if necessary.
\bibitem{nucooling}
Because it is almost non-interacting with nuclear matter, a neutrino 
tends to escape from the neutron star.  
This is the major cooling mechanism as the neutron star is being formed
in a supernova explosion.
George Gamow named this the URCA process, after a Brazilian 
casino where people lost a lot of money.
\bibitem{compressibility}
The reason for the ``(in)'' is because a materials physicist might rather
define compressibility as $\chi = -(1/V) (\partial V/ \partial p) = 
- (1/n) (d p/ dn)^{-1}$.
\bibitem{RHIC}
Folks interested in RHIC physics might want to, however. (RHIC stands for
``Relativistic Heavy Ion Collider,'' an accelerator at the Brookhaven
National Laboratory which is studying reactions like Au nuclei striking 
each other at center of mass energies around 200 GeV/nucleon.)
\end{thebibliography}
\end{document}